\documentclass[twocolumn,tradiabstract]{aa}
\usepackage{graphicx}
\usepackage{amsbsy}
\usepackage{natbib}
\usepackage{color}
\usepackage{txfonts}
\usepackage{url}
\usepackage{bm}

\newcommand{\vc}[1]{{\boldsymbol #1}}
\newcommand{\de}{\mathrm{d}}
\newcommand{\dpa}{\partial}

\DeclareMathSymbol{\varOmega}{\mathord}{letters}{"0A}
\DeclareMathSymbol{\varSigma}{\mathord}{letters}{"06}
\DeclareMathSymbol{\varPsi}{\mathord}{letters}{"09}
\DeclareMathSymbol{\varPhi}{\mathord}{letters}{"08}
\DeclareMathSymbol{\varGamma}{\mathord}{letters}{"00}

\newcommand{\Eq}[1]{Eq.~(\ref{#1})}

\newcommand{\Fig}[1]{Fig.~\ref{#1}}

\begin{document}

\title{Harvesting the decay energy of $^{26}$Al\\to drive lightning discharge in
protoplanetary discs}
\titlerunning{Lightning discharge in dense pebble filaments and particle discs}

\author{Anders Johansen\inst{1} \& Satoshi Okuzumi\inst{2}}
\authorrunning{Johansen}

\offprints{\\A.\ Johansen (\email{anders@astro.lu.se})}

\institute{$^1$ Lund Observatory, Department of Astronomy and Theoretical
Physics, Lund University, Box 43, 221 00 Lund, Sweden, \\e-mail:
\url{anders@astro.lu.se} \\ $^2$ Department of
Earth and Planetary Sciences, Tokyo Institute of Technology, Meguro, Tokyo,
152-8551, Japan}

\abstract{Chondrules in primitive meteorites likely formed by recrystallisation
of dust aggregates that were flash-heated to nearly complete melting. Chondrules
may represent the building blocks of rocky planetesimals and protoplanets in the
inner regions of protoplanetary discs, but the source of ubiquitous thermal
processing of their dust aggregate precursors remains elusive. Here we
demonstrate that escape of positrons released in the decay of the short-lived
radionuclide $^{26}$Al leads to a large-scale charging of dense pebble
structures, resulting in neutralisation by lightning discharge and flash-heating
of dust and pebbles. This charging mechanism is similar to a nuclear battery
where a radioactive source charges a capacitor. We show that the nuclear battery
effect operates in circumplanetesimal pebble discs. The extremely high pebble
densities in such discs are consistent with conditions during chondrule heating
inferred from the high abundance of sodium within chondrules. The sedimented
mid-plane layer of the protoplanetary disc may also be prone to charging by the
emission of positrons, if the mass density of small dust there is at least an
order of magnitude above the gas density.  Our results imply that the decay
energy of $^{26}$Al can be harvested to drive intense lightning activity in
protoplanetary discs.  The total energy stored in positron emission is
comparable to the energy needed to melt all solids in the protoplanetary disc.
The efficiency of transferring the positron energy to the electric field
nevertheless depends on the relatively unknown distribution and scale-dependence
of pebble density gradients in circumplanetesimal pebble discs and in the
protoplanetary disc mid-plane layer.}


\maketitle

\section{Introduction}

Tiny spherules that crystallised from molten rock are abundant in primitive
chondrite meteorites from the asteroid belt \citep{Krot+etal2009}. These {\it
chondrules} of typical sizes between 0.1 and 1 mm comprise approximately 60-80\%
of the mass of the ordinary and enstatite chondrite classes, mixed with smaller
metal grains and microscopic matrix particles \citep{Dodd1976}. The chondrule
abundance is more varied in the carbonaceous chondrites, ranging from
approximately 50\% in CO and CV chondrites to 0\% in the CI chondrites
\citep{ScottKrot2003}. The lack of chondrules in the CI chondrites could
nevertheless be the result of chondrule destruction during extensive aqueous
alteration of the parent body \citep{Endress+etal1996}.

The sheer abundance of chondrules implies that these particles played a key role
in the formation and/or growth of planetesimals in the solar protoplanetary
disc. Chondrule-sized particles can concentrate into dense clumps in the gaseous
protoplanetary disc either by preferential concentration between turbulent
eddies near the dissipation scale of the turbulent gas \citep{Cuzzi+etal2008} or
by the streaming instability that concentrates particles into dense filaments
through their collective drag force on the gas
\citep{YoudinGoodman2005,JohansenYoudin2007,BaiStone2010a}. The streaming
instability concentrates particles as small as chondrules at the location of
the asteroid belt if the turbulence in the protoplanetary disc is sufficiently
weak to allow mm-sized particles to sediment
\citep{Carrera+etal2015,YangJohansen2017}.  Chondrules also have ideal sizes to
become accreted onto young planetesimals and protoplanets.
\cite{Johansen+etal2015} showed that the observed size distribution of asteroids
(particularly the steeply declining differential size distribution for asteroids
larger than 100 km in diameter and the transition to a shallower size
distribution above 400 km in diameter) can arise due to the accretion of
chondrules with sizes similar to those measured in ordinary chondrites.
Chondrule accretion also drives the continued growth to Mars-sized protoplanets,
both in the asteroid belt and in the terrestrial planet region.  Therefore
understanding the origin of chondrules is imperative for understanding the
formation of asteroids and terrestrial planets.

The identification of radiogenic $^{26}$Mg in the CAI component
(calcium-aluminium-rich inclusions) of chondrites, produced through the decay of
the short-lived radionuclide $^{26}$Al \citep{Lee+etal1977}, provides a strong
connection between the solar protoplanetary disc and nearby supernovae in the
Sun's birth cluster. The ratio of $^{26}$Al to its stable sibling $^{27}$Al is
narrowly peaked at a value of approximately $5 \times 10^{-5}$ in the CAIs found
in the CV chondrites \citep{Amelin+etal2002,Connelly+etal2012}.  Age
determination of chondrules based on radiogenic $^{26}$Mg yields chondrule ages
that are systematically several million years younger than the CAIs
\citep{Villeneuve+etal2009}.  However, direct measurement of ages of individual
chondrules, based on the long-lived Pb-Pb system, indicates that the amount of
$^{26}$Al was originally lower in the chondrule-forming region compared to the
region where CAIs condensed out \citep{Connelly+etal2012,Larsen+etal2016}. This
could reflect a real heterogeneous distribution of $^{26}$Al in the solar
protoplanetary disc or that the dust grain carrier of $^{26}$Al was not
incorporated into chondrules (M.\ Bizzarro, personal communication, 2016). The
abundance of $^{26}$Al in the asteroid belt may have been as much as five times
lower than the value inferred from CAIs.

Decay of $^{26}$Al is believed to have heated some meteorite parent bodies to
differentiate completely while others (the parent bodies of the chondrites)
acquired an onion layering of internal heating degree below the melting
temperature. The measured systematic magnetisation of some chondrites
\citep{WeissElkins-Tanton2013}, supported by theoretical models of chondrule
accretion on existing planetesimals \citep{Johansen+etal2015}, indicates that
some differentiated bodies have outer layers of primitive chondrules. A
reduction in the primordial amount of $^{26}$Al in the asteroid belt bodies, as
indicated by the older Pb-Pb ages of chondrules \citep{Connelly+etal2012},
strongly reduces the degree of internal heating and hence the expected relative
frequency of fully differentiated, partially differentiated and primitive bodies
in the asteroid belt \citep{Larsen+etal2016}.

Chondrules appear to have crystallised after flash-heating of the unknown
chondrule precursor particles (likely fluffy dust aggregates of approximately mm
sizes). An additional component of small dust grains in meteorites, the matrix,
has been argued to have condensed out of a solar composition gas in connection
with chondrule formation, mainly based on the similar cooling times needed to
explain both chondrule and matrix crystal growth \citep{ScottKrot2005}. The
mechanism that led to widespread melting and partial evaporation of chondrule
precursors is one of the most troubling unsolved problems of meteoritics and
planet formation. Chondrule precursors must have been either heated and cooled
very quickly or heated in the close vicinity of other chondrule precursors, in
order to build up the saturation vapour pressure needed to stabilise the liquid
rock phase, a value much in excess of the ambient pressure of the solar
protoplanetary disc.  Heating a dust aggregate in the nominal pressure of the
solar protoplanetary disc will result in complete sublimation before the melting
temperature is reached, unless the heating and cooling were extremely rapid.
However, such rapid heating and cooling yield the wrong texture of the synthetic
chondrules produced in experiments \citep{Hewins+etal2005}.

The requirement for chondrule precursors to heat to the liquid phase without
vanishing by sublimation, combined with the observed lack of fractionation of
volatiles in chondrules, gives a number density constraint of minimum 10
chondrule precursors per m$^3$, or a minimum mass density of approximately
$10^{-5}$ kg m$^{-3}$ \citep{CuzziAlexander2006}. Retention of the volatile
element sodium during chondrule formation, inferred from the lack of internal
sodium gradients within large olivine crystals, implies even higher densities,
between $10^{-3}$ and $1$ kg m$^{-3}$ \citep{Alexander+etal2008}. The solar
protoplanetary disc had a characteristic density of $10^{-7}$ kg m$^{-3}$ at the
distance of the asteroid belt in its pristine stages and likely an order of
magnitude lower after a few million years of evolution \citep{Bitsch+etal2015}.
The Roche density -- above which particle clumps are gravitationally bound
against the tidal force from the central star -- is $\rho_{\rm R} \approx 2.8
\times 10^{-5}$ kg m$^{-3}$ at a distance of 2.5 AU from a solar-mass star.
Chondrules therefore appear to have heated in an environment of 100--10,000
times the Roche density.  \cite{AlexanderEbel2012} showed that matching the
abundance of compound chondrules, fusioned during cooling at the short-lived
viscous stage, also requires such ultrahigh densities. These conditions are hard
to reconcile with any nominal environment of the protoplanetary disc, except
perhaps very close to the star (0.1 - 0.5 AU) where the Roche density is 2--4
orders of magnitude higher than in the asteroid belt.

A plethora of theories have been formulated to explain the formation of
chondrules. The most popular is the shock model where chondrule precursors are
heated in shock waves that originate in the protoplanetary disc itself or from
planetesimals that plow through the gas on eccentric orbits. These models can
match the right heating and cooling times
\citep{DeschConnolly2002,Ciesla+etal2004}, but do not explain the high precursor
densities needed to sustain the liquid phase and maintain volatile abundances.
Current sheets driven by the magnetic field in the protoplanetary disc provide
another mechanism for heating chondrule precursors, but require a very high
starting temperature of more than 800 K to reach the relevant temperature for
melting chondrule  precursors \citep{McNally+etal2013}.

The planetesimal collision model posits that chondrules are secondary in nature
and arise as debris from planetesimal collisions. The collision model comes in
two flavours: collisions between pre-melted planetesimals
\citep{SandersTaylor2005} and collisions between solid planetary embryos
\citep{Johnson+etal2015}. The first actually taps into the decay energy of
$^{26}$Al, as the planetesimals melted through this energy input. The splash
ejecta is then envisioned to crystallise as mm-sized chondrules
\citep{Asphaug+etal2011}.  In the embryo model the melting happens in jets that
are ejected from the impact between solid embryos.  Collisions can in principle
explain the high densities of chondrule precursors as the collision debris
crystallises, as well as the retention of volatiles \citep{Dullemond+etal2014}.
Producing enough chondrules to form all the chondritic bodies that must have
existed in the primordial asteroid belt is nevertheless a challenge. In the
model of \cite{Johnson+etal2015}, creating even 1\% of the primordial asteroid
belt in chondrules required a total planetesimal mass in excess of three times
the mass of the primordial asteroid belt, as inferred from extrapolating between
the solid components of the terrestrial planets and the giant planets
\citep{Hayashi1981}. Most of the chondrules in the simulations of
\cite{Johnson+etal2015} were in fact produced in the terrestrial planet
formation region where collision frequencies are higher.

Lightning as a mechanism for heating chondrule precursors was proposed by
\cite{Whipple1966}. Lightning occurs in terrestrial volcanic ash plumes and
causes large ash particles to melt to form chondrule-like spherules
\citep{Genareau+etal2015}. Dust aggregates hit directly by an electric discharge
will nevertheless break rather than melt \citep{Guettler+etal2008}, but exposure
to the photons from the lightning results in melt spherules with size
distributions similar to chondrules \citep{EisenhourBuseck1995,Poppe+etal2010}.
In fact, terrestrial lightning processes approximately 70\% of its total energy
through photons \citep{Eisenhour+etal1994}. Radiation is attractive for
chondrule melting since it naturally introduces a smallest chondrule size, in
agreement with the observed log-normal or Weibull size distributions of
chondrules \citep{Dodd1976,Teitler+etal2010}, from the inefficient photon
absorption by small dust aggregates.

Obtaining lightning discharge in protoplanetary discs requires (a) a charging
mechanism, (b) a charge separation mechanism and (c) low enough gas conductivity
to allow the electric field to build up to a value where the current
collisionally ionises the hydrogen molecule. Charging and charge separation are
natural consequences of vertical or radial sedimentation, since large pebbles
are charged triboelectrically when colliding with dust grains
\citep{DeschCuzzi2000,Muranushi2010}. \cite{Pilipp+etal1998} nevertheless found
that vertical sedimentation of pebbles is many orders of magnitude too slow to
compete with the neutralising current. \cite{DeschCuzzi2000} considered charge
separation as chondrule-sized particles concentrate at km scales in strong disc
turbulence.  They found that the gas ionisation by decay of $^{26}$Al must be
reduced by several orders of magnitude in order to build up the break-down
electric field.

The elevated densities needed for chondrule formation make an intriguing link to
planetesimal formation by particle concentration by the streaming instability
and gravitational collapse \citep{Johansen+etal2014}. In this paper we therefore
explore lightning discharge and chondrule formation in dense pebble
environments. We show that secondary electrons released as positrons from the
decay of $^{26}$Al hammer through dust surfaces leads to positive charging of
the pebbles. Charge separation by radial drift is nevertheless very slow inside
dense pebble filaments and we find that the pebble current cannot compete with
the neutralising current, even if abundance of $^{26}$Al, and hence the gas
conductivity, is decreased by an order of magnitude. Therefore we consider
instead charge separation driven by the current of positrons released by the
decay of $^{26}$Al. Positrons stream out of dense pebble structures and cause a
large-scale negative charging of both circumplanetesimal discs and the
sedimented mid-plane layer of pebbles in the protoplanetary disc. This charging
is related to a direct-charging nuclear battery where a radioactive source
charges a capacitor. The electric field can build up values where the electrons
ionise the hydrogen molecule and cause lightning discharge. We construct simple
models of the circumplanetesimal pebble disc and the protoplanetary disc
environments. We show that the efficiency of converting the energy of the decay
of $^{26}$Al to photons released in lightning discharge can be high in both
these settings, although many uncertainties remain in the modelling of
circumplanetesimal discs and the charging of filamentary pebble structures with
density variation on many scales.

The paper is organised as follows. In Section \ref{s:problem} we review the
physics of lightning discharge and show by simple analytical arguments that
charge separation driven by differential drift of the pebbles relative to the
dust and gas is orders of magnitude slower than the neutralising current. This
conclusion is supported by thorough calculations of the equilibrium charge on
gas, dust and pebbles presented in Appendices \ref{s:charge}--\ref{s:Kdi}, where
we consider pebble charging by emission and absorption of positrons from the
decay of $^{26}$Al and release of secondary electrons after positron absorption
(this mechanism is elaborated on in Appendix \ref{s:secondary}).  In Section
\ref{s:nuclear_battery} we analyse the positron current emanating from pebble
filaments and sheets. We show that the positron current is higher than the
neutralising current when the dust-to-gas ratio is above approximately 50. We
discuss in Section \ref{s:application} the operation of the nuclear battery
effect in circumplanetesimal discs and in the sedimented mid-plane layer of
pebbles.  In the following Section \ref{s:heating} we demonstrate that the
cooling rates of chondrules formed in dense pebble regions agrees with the long
cooling times inferred from crystal growth experiments. Finally we summarise our
results in Section \ref{s:summary}.

\section{The problem with lightning discharge}
\label{s:problem}

In this section we review the physics of lightning discharge and demonstrate how
charge separation by pebble sedimentation or differential radial drift between
pebbles charged oppositely of dust grains and gas drives currents in
protoplanetary discs that are many orders of magnitudes lower than the
neutralising gas current.

\subsection{Lightning discharge}

Lightning discharge requires the build up of an electric field $E$ strong enough
that the accelerated electrons ionise the hydrogen molecule. This leads to a
cascade effect where the additional electrons increase the conductivity and
hence the charge separation is neutralised in thin discharge channels. The
breakdown electric field is given in \cite{DeschCuzzi2000} as
\begin{equation}
  E_{\rm BD} = 85 f_{\rm BD} \left( \frac{\rho_{\rm g}}{10^{-7}\,{\rm
  kg\,m^{-3}}} \right) \left( \frac{e_{\rm ion}}{60\,{\rm eV}} \right) \,{\rm
  V\,m^{-1}} \, .  \label{eq:EBD}
\end{equation}
Here we have normalised to a nominal gas density value $\rho_{\rm g}$ for the
asteroid belt region in the primordial state of the solar protoplanetary disc
and to an ionisation potential of $e_{\rm ion}=60$ eV. The ionisation energy of
H$_2$ is 15.6 eV, but \cite{DeschCuzzi2000} argued that impact energies below 60
eV lead to dissociation of the hydrogen molecule. \cite{OkuzumiInutsuka2015}
used a distribution function of electron energies and found that the breakdown
of the hydrogen molecule occurs already at electron energies around 5--10 eV;
hence we multiplied equation (\ref{eq:EBD}) by a factor $f_{\rm BD}$ that we
will assume in this paper to be 0.1.

The breakdown electric field can only be built up if the neutralising current,
driven by growing electric field, remains smaller than the current driven by the
motion of the charged solid particles. The gas has an intrinsic conductivity
that is a function of the density and mobility of ions and electrons. Ionisation
in the dense mid-plane layer is dominated by the decay of short-lived
radionuclides, most importantly by release of positrons in the decay of
$^{26}$Al \citep{DeschCuzzi2000}. External sources of ionisation, such as cosmic
rays, can be ignored in the dense mid-plane in the asteroid formation region.

\subsection{Ion and electron densities}

The gas conductivity depends strongly on the number densities of ions and
electrons, which are in turn dictated by adsorption onto dust grains. We
consider throughout this paper for simplicity a two-component model for the
solid particles. Component 1 consists of microscopic dust grains that are the
main absorbers of electrons and ions, while component 2 consists of macroscopic
(dust aggregate) pebbles that can form chondrules after flash heating. The
number densities of electrons, ions, positrons, small grains and pebbles are
$n_{\rm e}$, $n_{\rm i}$, $n_\beta$, $n_1$ and $n_2$. The two species of solid
particles have masses $m_1$ and $m_2$, radii $R_1$ and $R_2$, and charges (in
units of the elementary charge $e$) $q_1$ and $q_2$. The gas is characterised by
a neutral density $n_{\rm n}=\rho_{\rm g}/m_{\rm n}$, where $m_{\rm n}$ is the
neutral (hydrogen molecule) mass. Cross sections are denoted $\sigma$, with
single subscript used to indicate the target species or double subscript to
indicate that both collision partners matter for the cross section.

Positrons released by the decay of $^{26}$Al lose their energy to ionisation of
the hydrogen molecule (with ionisation potential $E_{\rm ion}$) and to stopping
in solid particles. The positron kinetic energy has an average value of
$T_\beta=0.66$ MeV and we take the energy loss $E_{\rm ion}=37$ eV per
ionisation \citep{Glassgold1995}, yielding in the absence of dust around 18,000
ionisations before the positron finally annihilates with a bound electron to
create a $\gamma$-ray.  This $\gamma$-ray, in turn, has a very long stopping
length and its effects on the mid-plane ionisation can be ignored. A more
typical end to the positron trajectory is penetration in a solid particle. The
stopping column density of positrons in silicon is $\varSigma_\beta \approx
3\,{\rm kg\,m^{-2}}$ and the stopping length is $\chi_{\rm s} \approx 1.3\,{\rm
mm}$ (see Section \ref{s:nuccurrent}). We use the terminology $R_1^*$, $R_2^*$,
$m_1^*$ and $m_2^*$ to denote the effective radii and masses that are available
for absorbing and releasing positrons. These quantities will only be different
from $R_1$, $R_2$, $m_1$ and $m_2$ if the particles are larger than $\chi_{\rm
s}$.  The effective radius can be written $R_i^*={\rm min}(R_i,\chi_{\rm s})$
and the effective mass $m_i^* = m_i - m_i (R<\chi_{\rm s})$, where we subtract
the inactive mass present deeper than $\chi_{\rm s}$ in the solid.

We can calculate the positron production rate per kilogram of rocky material
from the expression
\begin{equation}
  r_\beta = \frac{b x_{26}}{\tau m_{\rm H_2} Z_{\rm sil}} \, .
  \label{eq:rbeta}
\end{equation} Here $b=0.82$ is the branching ratio for positron emission,
$x_{26}=3\times10^{-10}$ is the nominal ratio of $^{26}$Al nuclei to hydrogen
molecules, $\tau$ is the decay time of $^{26}$Al (1.03 Myr) and $Z_{\rm sil}$ is
the dust mass fraction in rock \citep[we use the approximate solar
composition value of 0.005, following][]{Lodders2003}. This gives $r_\beta= 5
\times 10^{5} \, {\rm kg^{-1} s^{-1}}$. 

In regions of the protoplanetary disc where the number density of dust is very
high, the number densities of ions and electrons are given by a balance between
ionisation, at rate $\zeta$ per neutral molecule, and absorption by dust grains
\citep[e.g.][]{Okuzumi2009},
\begin{eqnarray}
  n_{\rm i} &=& \frac{\zeta n_{\rm n}}{n_d \sigma_d v_{\rm i}} \, ,
  \label{eq:ni} \\
  n_{\rm e} &=& \frac{\zeta n_{\rm n}}{n_d \sigma_d v_{\rm e}} \label{eq:ne} \,
  .
\end{eqnarray}
The index $d$ introduced here implies summation over dust species. The ion and
electron speeds, $v_{\rm i}$ and $v_{\rm e}$, are discussed further in Section
\ref{s:iecurrents}. The ionisation rate by positrons, $\zeta$, is given by the
expression
\begin{equation}
  \zeta = n_\beta \langle \sigma_{\rm n\beta} v_\beta \rangle \, .
  \label{eq:zeta}
\end{equation}
Here $\sigma_{\rm n\beta}$ is the ionisation cross section of molecular hydrogen
to positrons
\begin{equation}
  \sigma_{\rm n\beta} = \frac{m_{\rm n}}{\varSigma_\beta} \frac{T_\beta}{E_{\rm
  ion}} \approx 2.0 \times 10^{-23}\,{\rm m^2} \, ,
\end{equation}
where we have approximated the stopping column density in molecular hydrogen by
that in silicon, $\varSigma_\beta$.
The number density $n_\beta$ of positrons in the gas, moving at the speed
$v_\beta$, follows directly from the production equation
\begin{equation}
  \dot{n}_\beta = r_\beta \rho_{\rm p}^* - n_\beta n_d \sigma_d v_\beta
  (R_d^*/\chi_{\rm s}) - n_\beta n_{\rm n} \langle \sigma_{\rm n\beta} v_\beta
  \rangle (E_{\rm ion}/T_\beta) \, .
\end{equation}
The first term represents the emission rate, with $\rho_{\rm p}^* = n_1 m_1^* +
n_2 m_2^*$. The second term represents the absorption in solid particles and the
third term the absorption in gas molecules. The equilibrium state with
$\dot{n}_\beta=0$ yields
\begin{equation}
  n_\beta = \frac{r_\beta \rho_{\rm p}^*}{n_d \sigma_d v_\beta
  (R_d^*/\chi_{\rm s}) + n_{\rm n} \langle \sigma_{\rm n\beta} v_\beta
  \rangle (E_{\rm ion}/T_\beta)} \, .
  \label{eq:nbeta}
\end{equation}
We now rewrite the denominator in terms of the stopping column density
$\varSigma_\beta = \rho_\bullet \chi_{\rm s} = \rho_{\rm g} \chi_{\rm s,g}$,
where we introduced the material density $\rho_\bullet$ and the stopping length
in the gas $\chi_{\rm s,g}=(n_{\rm n} \sigma_{\rm n \beta})^{-1} (T_\beta/E_{\rm
ion})$ and assumed that the stopping column is independent of the material
(molecular hydrogen or silicate rock). That yields now, for solid particles with
radii much smaller than $\chi_{\rm s}$, the approximate expression
\begin{equation}
  n_\beta \approx \frac{r_\beta \rho_{\rm p}}{\rho_{\rm p}
  v_\beta/\varSigma_\beta + \rho_{\rm g} v_\beta/\varSigma_\beta} \, .
  \label{eq:nbetaapp}
\end{equation}
Combining equations (\ref{eq:nbetaapp}) and (\ref{eq:zeta}) yields
\begin{equation}
  \zeta = \zeta_0 \frac{Z/0.005}{1+Z}
  \label{eq:zeta2}
\end{equation}
with $\zeta_0=0.005 r_\beta \varSigma_\beta \sigma_{{\rm n}\beta} \approx 1.5
\times 10^{-19}\,{\rm s^{-1}}$ denoting the ionisation rate at the unsedimented
value $Z=0.005$. Here $Z=Z_1+Z_2$ is the total mass loading of solid particles
in the gas. The ionisation rate thus rises from $\zeta_0$ for unsedimented
particles well-mixed with the gas in the entire vertical column to $\zeta_\infty
= 200 \zeta_0 \approx 3 \times 10^{-17}\,{\rm s^{-1}}$ when the particle mass
loading is high, $Z\gg1$, e.g.\ in a dense mid-plane layer of sedimented
pebbles.

\subsection{Ion and electron currents}
\label{s:iecurrents}

Knowing the ion and electric densities from equation (\ref{eq:ni}) and
(\ref{eq:ne}) allows us to calculate the neutralising current of ions and
electrons as a function of the electric field strength $E$ through
\begin{eqnarray}
  \vc{J}_{\rm i} &=& -e n_{\rm i} \langle \vc{v}_{\rm i} \rangle \, ,
  \label{eq:Ji} \\
  \vc{J}_{\rm e} &=& +e n_{\rm e} \langle \vc{v}_{\rm e} \rangle \label{eq:Je}
  \, .
\end{eqnarray}
The mean ion and electron velocities, $\langle \vc{v}_{\rm i} \rangle$ and
$\langle \vc{v}_{\rm e} \rangle$  follow from the expression
\citep{MoriOkuzumi2016}
\begin{equation}
  \langle \vc{v}_\alpha \rangle = \frac{m_\alpha + m_{\rm n}}{m_\alpha m_{\rm
  n}} e \vc{E} \Delta t_\alpha \, .
  \label{eq:val}
\end{equation}
Here $\alpha$ denotes either ions or electrons. The mean time between collisions
is given for electrons as
\begin{equation}
  \Delta t_{\rm e} = \frac{1}{n_{\rm n} \langle \sigma_{\rm en} v_{\rm e}
  \rangle} 
\end{equation}
and for ions as
\begin{equation}
  \Delta t_{\rm i} = \frac{1}{1.6 \times 10^{-15}\,{\rm m^3\,s^{-1}} \times
  n_{\rm n}} \, .
\end{equation}
The constant in the denominator expresses that the interaction cross section
between ions and neutrals actually decreases with increasing ion speed, yielding
a constant product of cross section and speed \citep{Wannier1953}. The current
densities in equations (\ref{eq:Ji}) and (\ref{eq:Je}) are proportional to $E$
(through equation \ref{eq:val}) only if $n_{\rm i}$ and $n_{\rm e}$ are
independent of $E$. In fact, as first pointed out by \cite{OkuzumiInutsuka2015},
$n_{\rm i}$ and $n_{\rm e}$ can decrease with increasing $E$ when the electric
field is sufficiently strong.  If $E$ is above a threshold [see equations
(\ref{eq:Ecriti}) and (\ref{eq:Ecrit}) for the thresholds for ion and electron
heating, respectively], the electrons and ions obtain kinetic energies above the
mean thermal energy of the neutrals through acceleration by the electric field.
In the presence of such a strong electric field, the plasma particles collide
with and adsorb onto dust grains more frequently as the field strength
increases, and hence their number density decreases with increasing $E$.  For
the ion component, the added energy resides mainly in the mean motion because of
the larger inertia of the ion molecules (e.g., ${\rm HCO^+}$) than that of
neutrals (mainly ${\rm H_2}$ and He). Hence, for electrically heated ions, the
magnitude of the mean velocity $\langle \vc{v}_{\rm i} \rangle$ in equation
(\ref{eq:Ji}) becomes equal to the mean speed $\langle v_{\rm i} \rangle$ in
equation (\ref{eq:ni}).  Using this fact together with equations (\ref{eq:ni})
and (\ref{eq:Ji}), one can show that the neutralising ion current approaches the
constant expression \citep{OkuzumiInutsuka2015}
\begin{equation}
  J_{\rm i,\infty} = \frac{e \zeta n_{\rm n}}{n_d \sigma_d} \, .
\end{equation}
Inserting the ionisation rate from equation (\ref{eq:zeta2}), we find the
limiting values of the ion current, for low and high total particle mass
loading, as
\begin{eqnarray}
  J_{\rm i,\infty} &=& 1.9 \times 10^{-11}\,{\rm C\,m^{-2}\,s^{-1}} \times
  [(Z/Z_1)/10] \nonumber \\
  & & \times \left( \frac{R_1}{\rm \mu m} \right) \left(
  \frac{\rho_\bullet}{10^3\,{\rm kg\,m^{-3}}} \right) \quad {\rm for}\, Z \ll 1
  \label {eq:JiZl} \\
  J_{\rm i,\infty} &=& 1.9 \times 10^{-12}\,{\rm C\,m^{-2}\,s^{-1}}  \times
  Z_1^{-1} \nonumber \\
  & & \times \left( \frac{R_1}{\rm \mu m} \right) \left(
  \frac{\rho_\bullet}{10^3\,{\rm kg\,m^{-3}}} \right) \quad {\rm for}\, Z \gg 1
  \label {eq:JiZh}
\end{eqnarray}
We here assumed that the adsorption rate of ions is dominated by the dust
component, i.e.\ that adsorption onto pebbles can be ignored, and scaled the
result to dust grains of radius $R_1 = 1\,{\rm \mu m}$ and material density
$\rho_\bullet = 10^3\,{\rm kg\,m^{-3}}$. In equation (\ref{eq:JiZl}) we scaled
the ratio of the total mass loading of particles, $Z=Z_1+Z_2$, to the mass
loading in dust, $Z_1$, to 10. The ions experience significant heating, and
obtain their plateau current density value, above the critical electric field
strength of \citep{OkuzumiInutsuka2015}
\begin{eqnarray}
  E_{\rm crit,i} &=& \frac{m_{\rm i} m_{\rm n} \sqrt{3 k_{\rm B} T}}{(m_{\rm
  i}+m_{\rm n})^{3/2} e \Delta t_{\rm i}} \nonumber \\
  &\approx& 3.6 \times 10^{-1}\,{\rm V\,m^{-1}} \left( \frac{T}{180\,{\rm K}}
  \right)^{1/2} \left( \frac{\rho_{\rm g}}{10^{-7}\,{\rm kg\,m^{-3}}} \right) \,
  .
  \label{eq:Ecriti}
\end{eqnarray}
This critical field strength is 1-2 orders of magnitude smaller than the
breakdown strength (equation \ref{eq:EBD}) and has the same linear scaling with
gas density. Hence the ions are well in their heated regime at the electric
field strengths relevant for lightning discharge.

Electrons have larger inertia than neutrals and hence the electron kinetic
energy is dominated by the random motion. The critical electric field for
electron heating $E_{\rm crit}$ is \citep{OkuzumiInutsuka2015}
\begin{eqnarray}
  E_{\rm crit} &=& \sqrt{\frac{6 m_{\rm e}}{m_{\rm n}}} \frac{k_{\rm B} T}{e
  \ell_{\rm e}} \nonumber \\
  &\approx& 1.9 \times 10^{-3}\,{\rm V\,m^{-1}} \left(
  \frac{T}{180\,{\rm K}} \right) \left( \frac{\rho_{\rm g}}{10^{-7}\,{\rm
  kg\,m^{-3}}} \right) \, .
  \label{eq:Ecrit}
\end{eqnarray}
Here $\ell$ is the mean free path of the electrons to collide with neutrals.
The limiting values of the electron current at strong electric field
\begin{equation}
  J_{{\rm e},\infty} = \sqrt{\frac{\pi m_{\rm e}}{3 m_{\rm n}}} \frac{e \zeta
  n_{\rm n}}{\sigma_1 n_1} = \sqrt{\frac{\pi m_{\rm e}}{3 m_{\rm n}}} J_{\rm
  i,\infty} \approx 0.017 J_{\rm i,\infty} \, .
\end{equation}
Hence the electron current density plateaus at a value approximately 60 times
higher lower than the ion current density. We can therefore neglect the electron
current contribution to the neutralising current at the breakdown value of the
electric field.

\subsection{Sedimentation current}

Charge separation can be driven by aerodynamical size sorting, if small and
large solid particles are charged oppositely. Such opposite charging can e.g.\
be obtained by triboelectric charging in collisions between small dust and large
pebbles. \cite{DeschCuzzi2000} showed that triboelectric effects can charge
pebbles up to approximately $10^4$ electron charges. The charge separation in
particle components can then be transferred to a spatial charge separation by
the differential motion of the particles through the gas. Charged pebbles
sedimenting towards the mid-plane carry a current this way. The sedimentation
current is
\begin{equation}
  J_{\rm set} = n_2 v_2 q_2 \, .
\end{equation}
Here $n_2 = Z_2 \rho_{\rm g}/m_2$ is the pebble number density, $v_2$ is the
terminal speed of the pebbles and $q_2$ is the pebble charge. The terminal speed
at a height above the mid-plane of one scale-height, $z=H$, is given by
\begin{equation}
  v_2 = \frac{R_2 \rho_\bullet}{c_{\rm s} \rho_{\rm g}} \varOmega^2 H \, .
\end{equation}
Here $\varOmega$ denotes the Keplerian frequency at the distance $r$ from the
star and $\varOmega^2 H$ is the gravitational acceleration towards the
mid-plane. This expression simplifies to
\begin{eqnarray}
  v_2 &=& R_2
  \varOmega \frac{\rho_\bullet}{\rho_{\rm g}} = 0.5\,{\rm m\,s^{-1}}\, \left(
  \frac{R_2}{1\,{\rm mm}} \right) \left( \frac{\rho_\bullet}{10^3\,{\rm
  kg\,m^{-3}}} \right) \times \nonumber \\
  && \left( \frac{\rho_{\rm g}}{10^{-7}\,{\rm kg\,m^{-3}}} \right)^{-1} \left(
  \frac{r}{2.5\,{\rm AU}} \right)^{-3/2} \, .
\end{eqnarray}
That yields a sedimentation current density of
\begin{eqnarray}
  J_{\rm set} &=& \frac{3 Z_2 \varOmega q_2}{4 \pi R_2^2} = 9.7 \times
  10^{-20}\,{\rm C\,m^{-2}\,s^{-1}} \left( \frac{Z_2}{0.005} \right) \left(
  \frac{q_2}{10^4 e} \right) \times \nonumber \\
  && \left( \frac{R_2}{1\,{\rm mm}} \right)^{-2} \left( \frac{r}{2.5\,{\rm AU}}
  \right)^{-3/2} \, .
  \label{eq:Jset}
\end{eqnarray}
This is more than eight orders of magnitude lower than the neutralising ion
current for $Z_1=0.005$ (equation \ref{eq:JiZl}). The sedimentation current
would increase at a higher pebble mass loading $Z_2$, but this is not compatible
with the assumption that pebbles fall at the terminal speed at one scale-height
above the mid-plane. Increasing the pebble charge or decreasing the pebble size
and semi-major axis would also boost the current, but overall it seems hopeless
to gain the eight orders of magnitude needed to fight the neutralising gas
current.

\subsection{Radial drift current}

Particles also separate aerodynamically by radial drift caused by the radial
pressure support of the gas \citep{Weidenschilling1977,Nakagawa+etal1986}.
\cite{Tanaka+etal2005} provide general solutions for the equilibrium velocities
of gas and multiple particle components that couple via drag forces. The radial
velocity of pebbles relative to the smaller dust grains embedded in the gas
becomes
\begin{equation}
  v_{\rm dri} = \frac{2 {\rm St}_2}{Z_2} \Delta v \, .
  \label{eq:vdri}
\end{equation}
Here ${\rm St}_2 = R_2 \rho_\bullet/(H \rho_{\rm g})$ is the pebble Stokes
number and $\Delta v$ is the sub-Keplerian speed parameter
\citep{YoudinJohansen2007}. \cite{Tanaka+etal2005} introduced the inverse Stokes
number $\varGamma = {\rm St}^{-1}$ and we derived equation (\ref{eq:vdri}) in
the limit $\varGamma_1 \gg 1$, $\varGamma_2 \gg 2$, $\varGamma_1 \gg
\varGamma_2$ and $Z_2 \gg 1$. The relative drift speed at high $Z$ is dominated
by the outwards motion of gas and dust pushed through the almost stationary
pebble component, but given charge neutrality $q_2 n_2 = -(q_1 n_1 + n_{\rm i} -
n_{\rm e})$ we can use $q_2 n_2$ to describe the charge density of the current.
Inserting typical parameters for the pebbles and the gas we obtain
\begin{eqnarray}
  v_{\rm dri} &=& 0.064\,{\rm m\,s^{-1}}
  Z_2^{-1}
  \left( \frac{R_2}{1\,{\rm mm}} \right) \times \left(
  \frac{\rho_\bullet}{10^{-3}\,{\rm kg\,m^{-3}}} \right) \times \nonumber \\
  &&
  \left( \frac{\rho_{\rm
  g}}{10^{-7}\,{\rm kg\,m^{-3}}} \right)^{-1} \left( \frac{\Delta v}{50\,{\rm
  m\,s^{-1}}} \right) \left( \frac{r}{2.5\,{\rm AU}} \right)^{5/4} \, .
\end{eqnarray}
The low mass loading limit ($Z_2 \ll 1$) yields the same value and scaling, but
without the dependence on $Z_2$. The drift current density for high mass loading
is then
\begin{eqnarray}
  J_{\rm dri} &=& \frac{3 \Delta v q_2}{2 \pi R_2^2 H} \approx
  2.5 \times 10^{-18}\,{\rm C\,m^{-2}\,s^{-1}} \times \left(
  \frac{q_2}{10^4 e} \right) \times \nonumber \\
  & & \left( \frac{\Delta v}{50\,{\rm m\,s^{-1}}} \right) \left( \frac{R_2}{\rm
  mm} \right)^{-2} \left( \frac{r}{2.5\,{\rm AU}} \right)^{5/4} \, .
  \label{eq:Jdri}
\end{eqnarray}
The maximum radial drift current is thus more than 20 times higher than the
maximum sedimentation current [equation (\ref{eq:Jset})] at $Z_2 \gg 1$, but
only becomes comparable to the neutralising gas current at the breakdown
electric field [equation (\ref{eq:JiZh})] for $Z_1 \sim 10^6$.

\subsection{Full model}

The derivations presented in the previous subsections were made under a set of
simplifying assumptions, e.g.\ that the adsorption rate of electrons and ions on
dust particles are not affected by the charge of the dust. We also rather
handwavingly referred to the triboelectric charging model of
\cite{DeschCuzzi2000} when setting the characteristic charge of pebbles to $q_2
= 10^4 e$. In Appendix \ref{s:charge} and Appendix \ref{s:equilibrium} we
therefore present a fully numerical solution to the complex charge equation
equilibrium. We include the release of secondary electrons from penetrating
positrons and show that this leads to a large positive charging of the pebbles,
at the expense of a negative charge on the gas. However, the results confirm
that aerodynamical charge separation is much too slow to compete with the
neutralising current.  Therefore we consider in the next section a current
driven by the positrons themselves.

\section{Nuclear battery effect}
\label{s:nuclear_battery}

We now describe a charging mechanism akin to a nuclear battery. The simplest
form of a direct charging nuclear battery has a radioactive isotope on one plate
of a capacitor. The nuclear decay products are absorbed by the opposite plate.
If the emitted particles are charged (positrons, electrons, alpha particles or
fission fragments), then they will charge up both plates and build up a voltage
difference over the capacitor. In a protoplanetary disc, dense pebble structures
emit positrons from the decay of $^{26}$Al. This leads to a net negative
charging of the pebbles -- provided that the positrons stream into regions of
significant gas that stops the positrons but does not emit any positrons in the
opposite direction. Importantly, there will only be a net current if the
positrons are stopped by gas, as the dust component is in itself in complete
balance between emitting and absorbing positrons.

\subsection{Characteristic current}
\label{s:nuccurrent}

The characteristic positron current density arising from a pebble region, dense
enough to be optically thick to its own positrons, is
\begin{equation}
  J_\beta = \frac{1}{6} e r_\beta \rho_{\rm p} \ell_\beta \, .
\end{equation}
Here $e$ is the positron charge, $r_\beta$ is the emission rate of positrons
per unit dust mass and unit time, $\rho_{\rm p}$ is the local mass density of
solids and $\ell_\beta$ is the stopping length of the positrons. We divided the
expression by a factor six to take into account the random direction of the
positrons. We can express the stopping length in terms of the stopping
column $\varSigma_\beta = \rho_{\rm p} \ell_\beta$, yielding the positron
current expression
\begin{equation}
  J_\beta = \frac{1}{6} e r_\beta \varSigma_\beta \approx 3.6 \times
  10^{-14}\,{\rm C\,m^{-2}\,s^{-1}} \times f_{26} \, .
  \label{eq:Jpos}
\end{equation}
where the activity of $^{26}$Al relatively to the value at the formation
conditions of CAIs, $f_{26}$, is defined as
\begin{equation}
  f_{26} = \frac{r_\beta}{5 \times 10^5\,{\rm kg^{-1} s^{-1}}} \, .
\end{equation}
We used here $\varSigma_\beta \approx 3$ kg/m$^2$ in silicon\footnote{We
calculated the stopping length $\ell_\beta \approx 1.3$ mm for 0.66 MeV
positrons and internal density $\rho_\bullet=2.3 \times 10^3$ kg m$^{-3}$, using
the positron penetration calculations presented in Appendix \ref{s:secondary};
see also \cite{Umebayashi+etal2013} who obtained a similar value of the positron
stopping length.}. This current density is orders of magnitude higher than the
currents carried by pebble sedimentation [equation (\ref{eq:Jset})] or radial
drift [equation (\ref{eq:Jdri})] and is larger than the neutralising ion current
for dust mass-loadings $Z_1 > 50$ [equation (\ref{eq:JiZh})]. Such mass loading
of small dust in the gas is hard to achieve under nominal conditions in the
protoplanetary disc. However, in Section \ref{s:application} we apply the
nuclear battery effect to dense filaments in this mid-plane layer and to
circumplanetesimal pebble discs. Dust is produced continuously in collisions
between pebbles and hence the dust density is expected to follow the density of
larger pebbles. In the coagulation-fragmentation simulations of
\cite{Birnstiel+etal2011} small dust constitutes approximately 1-10\% of the
mass in pebbles of mm-cm sizes. We generally assume in this paper that this dust
fraction is valid also in overdense pebble regions. The pebble-pebble collision
timescale can be estimated as 
\begin{eqnarray}
  t_{\rm coll} &=& ( 4 \pi R_2^2 n_2 \delta v )^{-1} \nonumber \\
  &\approx& 3 \times 10^3\,{\rm s}\, \left( \frac{R_2}{\rm mm} \right) \left(
  \frac{\rho_2}{10^{-2}\,{\rm kg\,m^{-3}}} \right)^{-1} \left( \frac{\delta
  v}{0.1\,{\rm m\,s^{-1}}} \right)^{-1} \, .
  \label{eq:tcoll}
\end{eqnarray}
The first lightning discharge will thus be delayed until small dust has been
produced to reduce the conductivity. However, the dust production time-scale to
produce small dust fragments is much shorter than the decay time of $^{26}$Al,
within the range of likely physical parameters in equation (\ref{eq:tcoll}). We
assume that the dust within dense pebble filaments undergoes continuous release
in pebble-pebble collisions and recoagulation to maintain a high dust mass
loading for millions of years.
  
As we discuss further in the next subsection, the neutralising gas current in
dense pebble structures may be even lower than the expression in equation
(\ref{eq:JiZh}), if the gradient in the particle density is high. In that case
positrons streaming out of the region are not replaced by positrons streaming in
and hence the ionisation rate inside of the structure is reduced.

\subsection{Optimal conditions for the nuclear battery}

Consider a dense pebble region emitting positrons that are absorbed in the
surrounding gas. The charging of the pebbles builds up an electric field that
points from the surroundings into the negatively charged pebble region.
Initially the positrons are stopped by collisions with gas molecules, but as the
pebble region becomes increasingly negatively charged, the positrons lose more
and more of their kinetic energy to work against the electric field. The width
of the pebble structure is now key to obtaining neutralisation through lightning
discharge. If the pebble structure has a too large width, $W$, then the
positrons are trapped in the pebble region before the breakdown electric field
value is reached. The critical width $W^\dagger$ is given through the relation
\begin{equation}
  T_\beta = e W^\dagger E_{\rm BD} \, .
\end{equation}
Here $T_\beta=0.66\,{\rm MeV}$ is the kinetic energy of the positrons, $e$ is
the positron charge and $E_{\rm BD}$ is the breakdown value of the electric
field. Using the expression for the breakdown electric field from equation
(\ref{eq:EBD}) we obtain
\begin{equation}
  W^\dagger = 66\,{\rm km}\,\left( \frac{\rho_{\rm g}}{10^{-7}\,{\rm kg\,m^{-3}}}
  \right)^{-1} \left( \frac{f_{\rm BD}}{0.1} \right)^{-1} \, .
  \label{eq:width}
\end{equation}
Here $f_{\rm BD}$ defines the ratio of the actual breakdown field to the nominal
value given in \cite{DeschCuzzi2000}. A one-dimensional pebble filament should
have a width comparable to $W$ for efficient conversion of positron energy into
lightning discharge; broader filaments do not reach the breakdown field strength
at all and narrower filaments reach the breakdown field strength over such short
length-scales that positrons only deposit a minor fraction of their energy in
the electric field. Two-dimensional pebble sheets, on the other hand, build up
an electric field that has a constant value up to a height comparable to the
extent of the sheet itself. Hence a pebble sheet loads energy into the electric
field efficiently independently of the width, for any width less than
$W^\dagger$. Reaching the breakdown electric field, ions and electrons
accelerated by the electric field to neutralise the pebble region must now
release an energy similar to the original kinetic energy of the positrons. This
way the energy of the positron emission becomes converted to kinetic energy and
heat in the lightning discharge current.

The density of the pebble structure does not affect the critical width
found in equation (\ref{eq:width}). However, the highest positron current is
obtained when the column density of the pebble structure is similar to the
stopping column of the positrons. Structures of lower column density exhibit a
lower current density than the characteristic value given in equation
(\ref{eq:Jpos}), while structures of higher column density self-absorb most of
the emitted positrons and therefore have a low efficiency in converting the
positron emission to charge separation. The highest positron current combined
with the highest charging efficiency are therefore obtained when the pebble
density obeys
\begin{equation}
  \rho_{\rm p} W = \varSigma_\beta \, .
\end{equation}
That gives a pebble density of
\begin{equation}
  \rho_{\rm p} = \frac{\varSigma_\beta}{W} = 5 \times 10^{-5}\,{\rm
  kg\,m^{-3}}\, \left( \frac{W}{66\,{\rm km}} \right)^{-1} \, .
\end{equation}
Higher pebble densities yield inefficient charging and lower pebble
densities result in a reduction of the positron current. Combining with the
critical width for conversion of positron kinetic energy to electric
potential, from equation (\ref{eq:width}), yields the optimal solids-to-gas
ratio for a pebble filament and the minimum solids-to-gas ratio for a pebble
sheet as
\begin{equation}
   \frac{\rho_{\rm p}}{\rho_{\rm g}} \approx 500 \frac{f_{\rm BD}}{0.1} \, .
\end{equation}
Lower solids-to-gas ratios can still yield efficient conversion of positron
energy to heating, but the
current emanating from pebble structures of lower densities will be lower than
the characteristic current. The characteristic current is nevertheless quite
high, at $J_\beta=3.6\times10^{-14}\,{\rm C\,m^{-2}\,s^{-1}}$, so even a
factor 10 reduction yields currents in excess of the neutralising current at
$Z_1=500$.
\begin{figure}
  \begin{center}
    \includegraphics[width=\linewidth]{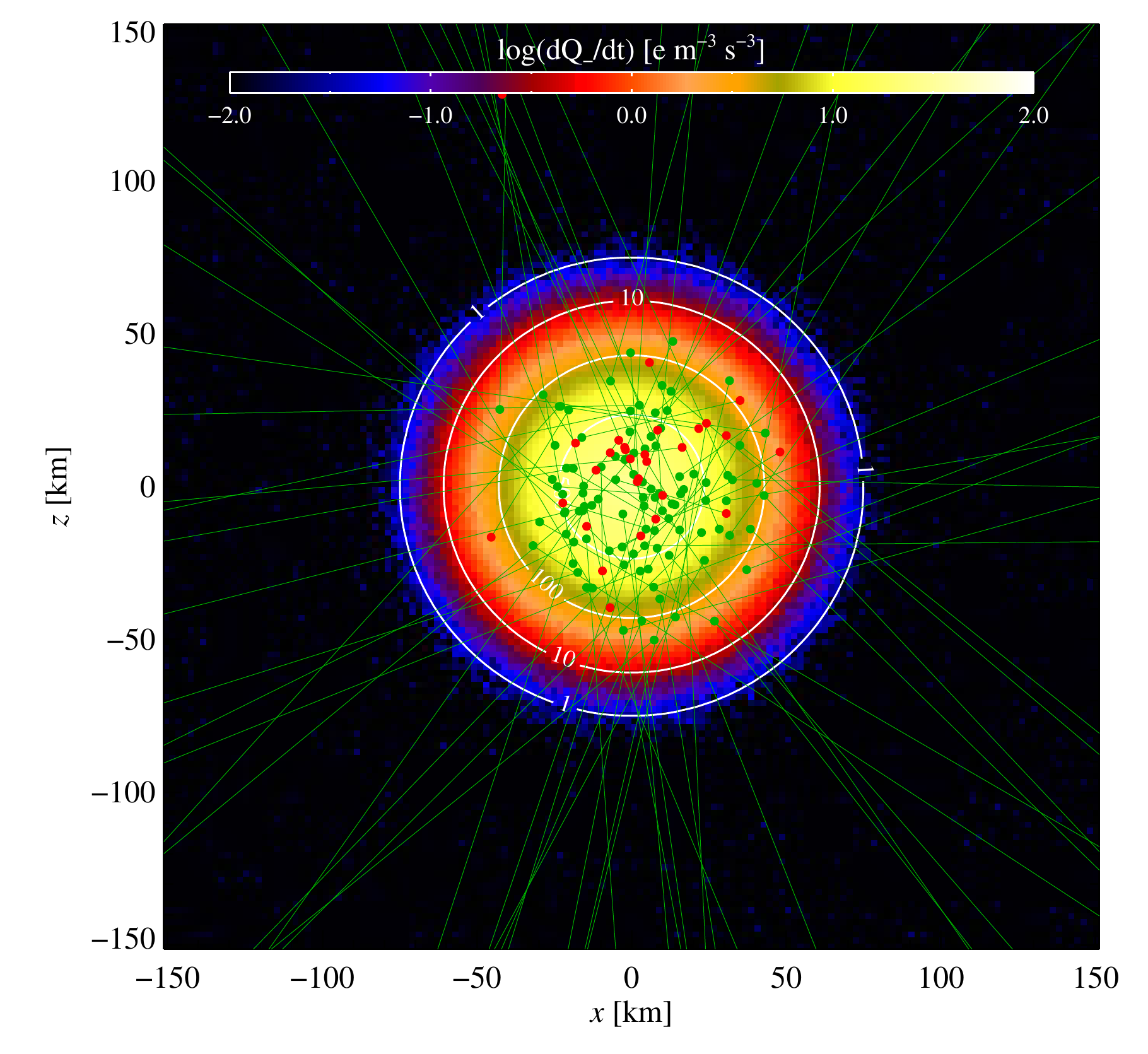}
  \end{center}
  \caption{The charging rate by positron release in an axisymmetric pebble
  filament of half-width 20 km. The filament has a peak density of 1000 times
  the gas density and the background dust density is 0.1 times the gas density.
  The colored contours show the negative charging rate and the white contour
  lines show the dust-to-gas ratio. The charging rate has been obtained by a
  Monte Carlo calculation of the emission and absorption of $10^7$ positrons. A
  hundred randomly picked positron trajectories are shown in green, released at
  the green dot and absorbed at the red dot (which happens mostly outside of the
  plotted region). The filament charges rapidly negative as most of the released
  positrons are absorbed in the surrounding gas.}
  \label{f:Qdot_rhop}
\end{figure}

\subsection{Simulations of nuclear battery effect}
\begin{figure}
  \begin{center}
    \includegraphics[width=\linewidth]{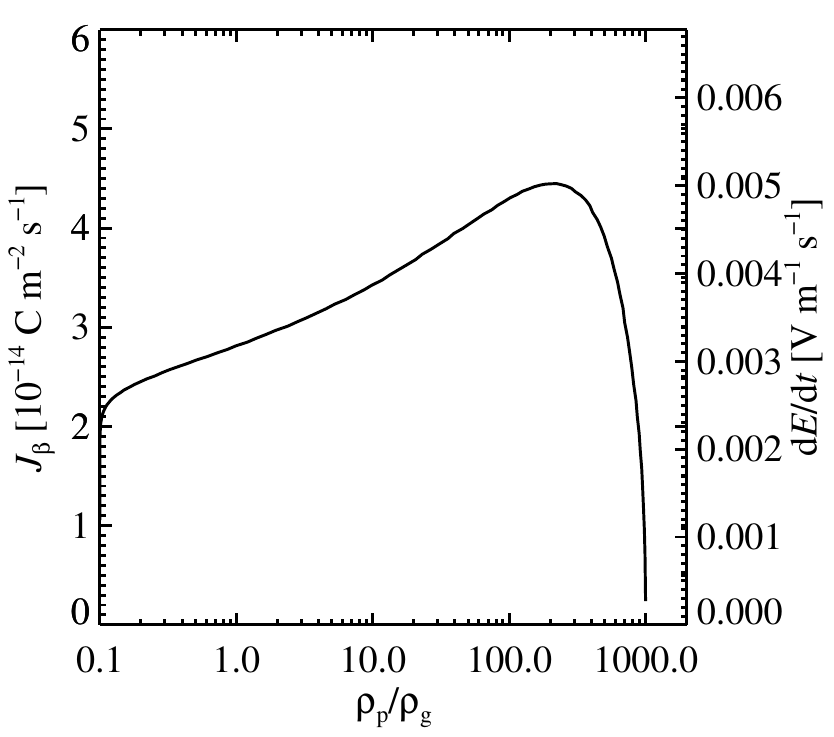}
  \end{center}
  \caption{The current density of the positrons as a function of the
  solids-to-gas ratio in the particle filament shown in Figure
  \ref{f:Qdot_rhop}. The current density is above $2 \times 10^{-14}\,{\rm
  C\,m^{-2}\,s^{-1}}$ through the entire filament. The right axis shows the
  rate of change of the electric field. The breakdown electric field will be
  reached in the filament after a characteristic time-scale of only
  approximately $10^3$ s.}
  \label{f:Jp_epsp}
\end{figure}
The calculations presented in the previous subsection do not capture the details
of the positron emission and absorption process. To get a better handle on
realistic values of the positron current and the charging efficiency we perform
a Monte Carlo simulation of positron emission and absorption in a dense particle
filament. We use a Gaussian profile for the density of particles, with peak mass
loading $\epsilon_2$ relative to the gas, embedded in a constant background
particle density with mass loading $\epsilon_1$,
\begin{equation}
  \rho_{\rm p}=\epsilon_1 \rho_{\rm g}+\epsilon_2 \rho_{\rm g}
  \exp[-(x^2+z^2)/(2 W^2)] \, .
\end{equation}
Note that $\epsilon_1$ and $\epsilon_2$ do not represent respectively small and
large particles (and hence are unrelated to $Z_1$ and $Z_2$ employed in Section
\ref{s:problem}); from the view of positron emission the particle size is
insignificant, as long as the particles are smaller than approximately 1 mm, so
that the positron can freely escape from the interior. We create positrons at
random positions that are weighted with the ratio of the local particle density
$\epsilon_1+\epsilon_2$ to the total dust mass. The positrons are given random
3-D directions and are absorbed after interacting with the
characteristic stopping column of $\varSigma_\beta$ in gas and particles. In
Figure \ref{f:Qdot_rhop} we present the results of a simulation with $\rho_{\rm
g}=10^{-7}\,{\rm kg\,m^{-3}}$,
$\epsilon_1=0.1$, $\epsilon_2=1000$ and $W=20$ km. The center of the filament
charges rapidly negative, at 100 electron charges per cubic meter per second.
The measured current density streaming out of the filament is shown in Figure
\ref{f:Jp_epsp}, as a function of the solids-to-gas ratio (a measure of the
distance from the centre of the filament). The current density is above $3
\times 10^{-14}$ C m$^{-2}$ s$^{-1}$ in most of the region and peaks at a value
of $4.5 \times 10^{-14}$ C m$^{-2}$ s$^{-1}$, near the characteristic value
found in equation (\ref{eq:Jpos}). The rate of change of the electric field is
related to the current density through the simple expression $\dot{E} =
\epsilon_0^{-1} J$, with vacuum permittivity $\epsilon_0$. The right axis of
Figure \ref{f:Jp_epsp} shows that the breakdown value of the electric field is
reached within a characteristic time-scale of only $10^3$ s. The filament thus
quickly builds up an electric field strength near the breakdown value and
the continued charging by positron release is neutralised by lightning, keeping
the electric field around the breakdown value.

The efficiency of the nuclear battery effect can be calculated from the net
charging. We perform here a series of experiments where we cover a pebble
structure and its surroundings by a fixed grid and monitor the emission and
absorption of positrons in each cell. We consider a gas density of $\rho_{\rm g}
= 10^{-7}\,{\rm kg\,m^{-3}}$ and a box size of $10^5$ km to capture the positron
stopping length in the gas of approximately $3 \times 10^4$ km. The dust-to-gas
ratio was chosen to be $\epsilon_1=0.1$ and the pebble-to-gas ratio varied from
$\epsilon_2=1$ to $\epsilon_2=10^3$. The pebble structure scale is fixed at $3
\times 10^3$ km, larger than the example shown in in Figure \ref{f:Qdot_rhop}
because we must be careful to capture here the large length-scales of positron
absorption in the gas. The transition between being thin and thick to positron
emission happens then at a pebble mass loading of $\epsilon_2 \approx 10$.

\begin{figure}
  \begin{center}
    \includegraphics[width=0.9\linewidth]{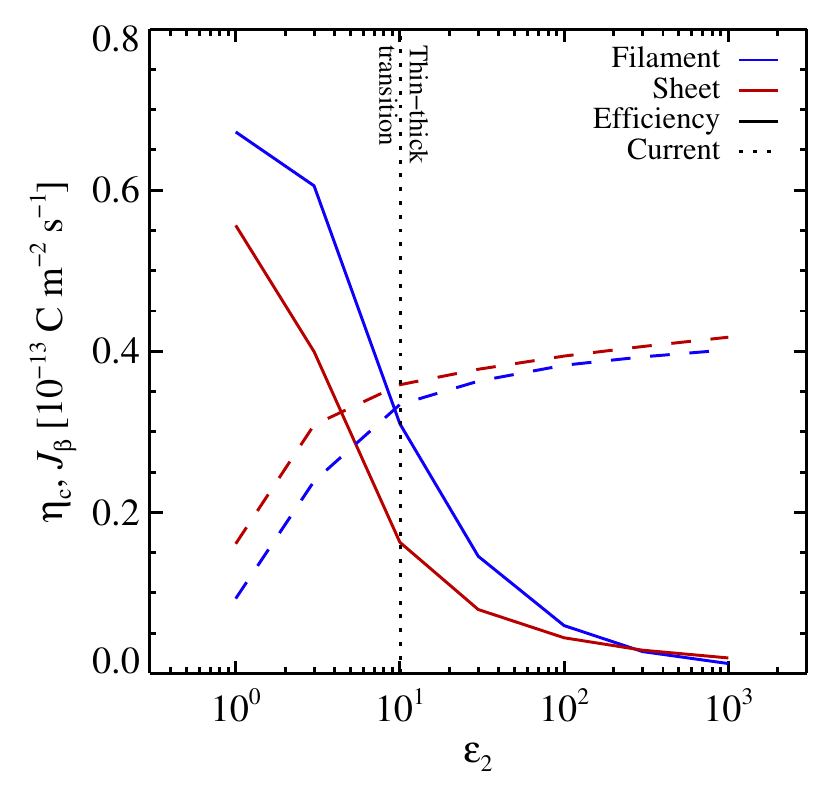}
  \end{center}
  \caption{Charging efficiency (solid lines) and current density (dashed lines)
  as a function of the mass loading in pebbles, for both 1-D filaments (blue
  lines) and 2-D sheets (red lines), for a characteristic structure width of
  3,000 km. The charging efficiency is high at low pebble densities when the
  positrons can escape freely from the structure, but falls for high pebble
  densities due to self-absorption of positrons within the structure. The
  current behaves oppositely: at low pebble densities the current is low, since
  the structure is thin to its own positron emission, while the current density
  plateaus around the characteristic value of $3.6 \times 10^{-14}\,{\rm
  C\,m^{-2}\,s^{-1}}$ for high pebble densities.}
  \label{f:efficiency}
\end{figure}

Releasing a total of $N_\beta$ positrons inside the pebble structure and
measuring a net charge of $Q_j$ on the grid, where $j$ is the grid index, yields
an efficiency of
\begin{equation}
  \eta_{\rm c} = \frac{\sum_j |Q_j|}{N_\beta} \, .
\end{equation}
Here we sum only over regions inside of the pebble structure, defined as $\rho_2
> \epsilon_1 \rho_{\rm g}$. This exclusion is necessary since positrons released
near the boundary of the computational domain are not replaced with a
counterstreaming flux and hence appear to have high charging efficiency. A
uniform particle density would yield $\eta_{\rm c} \approx 0$, since all cells
would have equally many emitted and absorbed positrons. The efficiency of the
nuclear battery effect is shown in Figure \ref{f:efficiency} as a function the
pebble mass loading $\epsilon_2$. We consider both 1-D filaments and 2-D sheets.
The most efficient charging is obtained when the pebble density is low, as most
positrons can escape such dilute pebble structures. The current is only
$10^{-14}$ C m$^{-2}$ s$^{-1}$ for low values of the pebble density, but
plateaus around the characteristic value of $3.6  \times 10^{-14}\,{\rm
C\,m^{-2}\,s^{-1}}$ when the pebble structure transitions from being thin to
thick to its own positron emission (at $\epsilon_2 \approx 10$).

\section{Application to protoplanetary discs}
\label{s:application}

The nuclear battery effect identified and described in Section
\ref{s:nuclear_battery} provides a pathway to convert the decay energy of
$^{26}$Al to lightning discharge that can flash heat solids. The total energy
present in positrons released by $^{26}$Al -- of decay rate
$r_\beta=5\times10^5\,{\rm kg^{-1}\,s^{-1}}$, energy $T_\beta = 0.66$ MeV and
decay constant $\tau_{26} = 1.03$ Myr -- is
\begin{equation}
  E_{26} = r_\beta T_\beta \tau_{26} \approx 1.7 \times 10^6\,{\rm J\,kg^{-1}}
  \, .
\end{equation}
The heating and partial melting of chondrule precursors requires approximately
$1.6 \times 10^6\,{\rm J\,kg^{-1}}$ \citep{HeveySanders2006}. The positrons thus
release, over the life-time of $^{26}$Al of a few million years, approximately
the energy needed to melt all rock in the protoplanetary disc\footnote{Note that
the full heating capacity of $^{26}$Al trapped inside a planetesimal is much
higher, at 4 MeV per decay compared to 0.66 MeV that goes into the kinetic
energy of the positron, due to eventual annihilation of the positron to produce
a $\gamma$-ray.}. Transferring the positron energy to the electric field and
then to lightning discharge can therefore lead to flash heating of a significant
fraction of the solids in the disc, if the energy transfer and heating
efficiencies are both high. We discuss here two settings where the nuclear
battery effect could operate to thermally process solids: in pebble discs
orbiting young planetesimals and in the sedimented mid-plane layer of pebbles in
the main protoplanetary disc.

\subsection{Nuclear battery in a circumplanetesimal pebble disc}

The high abundance of sodium within chondrules can be understood if chondrule
precursors were heated in an environment that was several orders of magnitude
denser than the Roche density at the distance of the asteroid belt from the Sun
\citep{Alexander+etal2008}. Such dense regions should be undergoing
gravitational collapse. The gravitational collapse phase is short, on the order
of a few orbital periods \citep{Nesvorny+etal2010,WahlbergJanssonJohansen2014}.
How can chondrules in chondrites reflect formation conditions that were
prevalent for only the last tiny fraction of the free life of a pebble in a
protoplanetary disc before it was incorporated in a planetesimal? Actually,
densities higher the Roche limit do not necessarily imply gravitational
collapse, if the collapsing region is rotationally supported.
\cite{JohansenLacerda2010} demonstrated how pebbles accreted onto protoplanets
enter a prograde particle disc in stable orbit around the central protoplanet.
The Roche density in such a particle disc will be orders of magnitude higher
than the Roche density that results from the tidal force of the central star
alone. Could the high particle densities required for chondrule heating have
been sustained for hundreds of thousands of years in thin pebble discs orbiting
around planetesimals and protoplanets?
\begin{figure*}
  \begin{center}
    \includegraphics[width=0.8\linewidth]{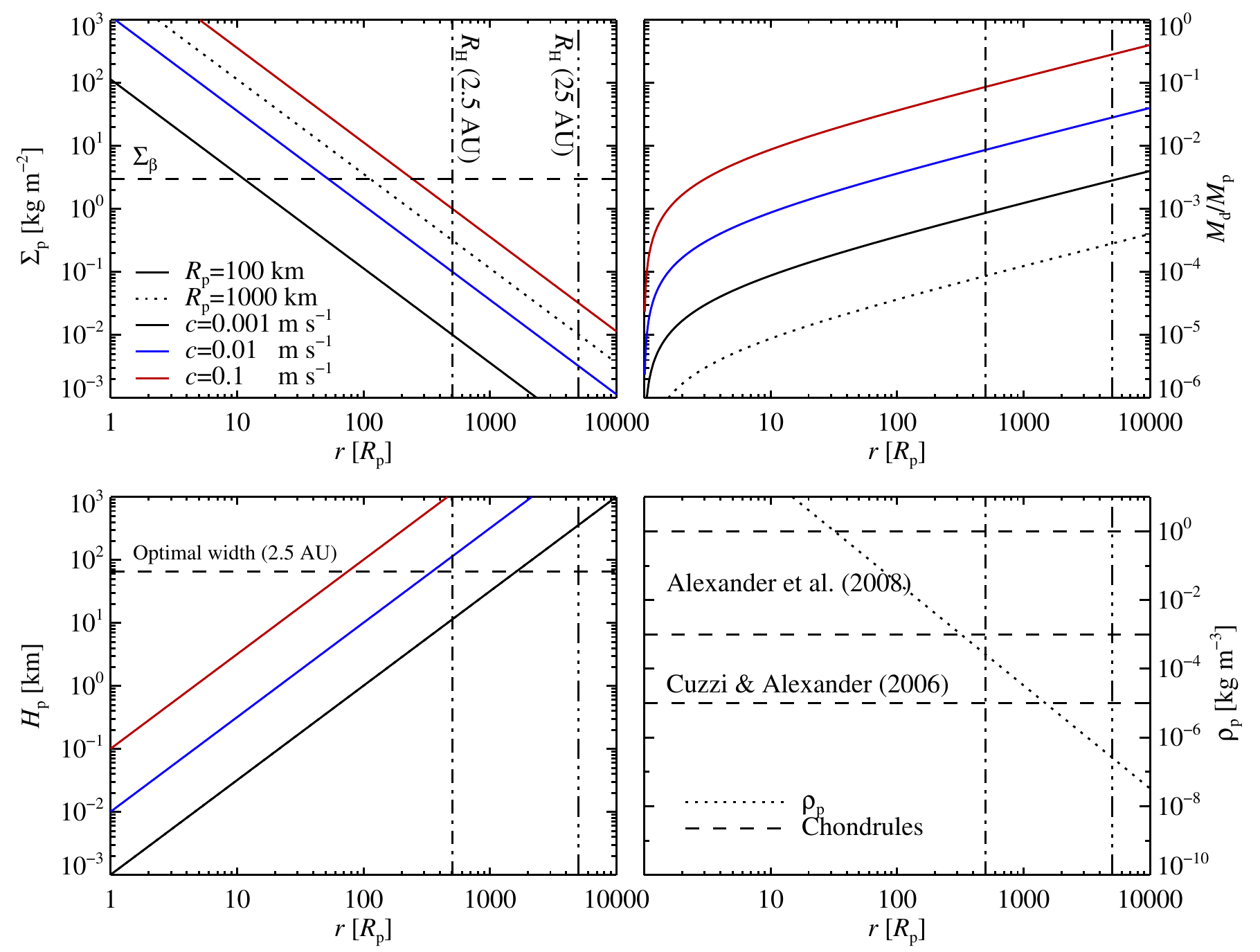}
  \end{center}
  \caption{The conditions for the nuclear battery effect in circumplanetesimal
  discs. The {\it top left} plot shows the pebble column density as a function
  of the distance from the planetesimal, for two values of the planetesimal
  radius ($R_{\rm p}=100\,{\rm km}$ and $R_{\rm p}=1000\,{\rm km}$) and three
  values of the random pebble speed $c$. The vertical dot-dashed lines indicate
  the
  Hill radius at 2.5 AU and 25 AU, respectively, and the horizontal dashed line
  indicates the stopping column of positrons. The pebble column density is below
  the stopping column of positrons in the outer region of the circumplanetesimal
  disc, unless the random pebble motion is very high, so positron emission
  leads to an efficient charging of the circumplanetesimal disc.  The {\it top
  right} plot shows the integrated disc mass relative to the mass of the central
  planetesimal. The small planetesimal has a relatively more massive disc than
  the larger protoplanet and hence can convert a higher fraction of the central
  mass to chondrules. The {\it bottom left} plot shows the scale-height of the
  pebble disc. The stopping length of positrons at the breakdown value of the
  electric field is indicated for the nominal breakdown current at 2.5 AU. The
  {\it bottom right} plot shows the mid-plane density of pebbles compared to the
  minimum pebble density required for lack of isotopic fractionation and
  stability of the liquid phase by \cite{CuzziAlexander2006} and the density
  range implied by the observed high abundance of Na within chondrules by
  \cite{Alexander+etal2008}. The pebble density in the circumplanetesimal disc
  environment lies perfectly within these experimental constraints.}
  \label{f:planetesimal_lightning}
\end{figure*}

A circumplanetesimal particle disc will not evolve in isolation from its
surroundings. The sub-Keplerian gas acts as a friction on the outermost
pebbles, unbinding them from the gravity of the planetesimal so that they mix
with other pebbles in the protoplanetary disc and participate in new
generations of planetesimal formation and pebble accretion. Planetesimals that
enter eccentric orbits, e.g.\ perturbed by the growing protoplanets
\citep{Johansen+etal2015}, experience even stronger gas headwinds that could
erode off pebbles even more efficiently. A large fraction of the chondrules
that are flash-heated in a circumplanetesimal disc are nevertheless expected to
become accreted onto the central planetesimal. Radiometric dating of individual
chondrules, by the long-lived Pb-Pb chronometer, shows that chondrules in the
same chondrite display a wide range of ages \citep{Connelly+etal2012}. This age
difference could reflect either that the parent bodies of those meteorites
accumulated late from a mixture of chondrules that formed in other
circumplanetesimal discs or that the chondrules that are accreted through a
circumplanetesimal disc experience efficient mixing between chondrules formed
at various epochs.

We can estimate the column density of a circumplanetesimal disc from the
assumption that the disc evolves towards a state that is marginally unstable
to self-gravity. In this view the circumplanetesimal disc accretes mass onto
the central planetesimal by large-scale gravitational stresses when onfalling
pebbles lower the Safronov-Toomre $Q$ parameter to below unity
\citep{Safronov1960,Toomre1964}. The marginal case of $Q=1$ gives a column
density of
\begin{equation}
  \varSigma_{\rm p} = \frac{c \varOmega_{\rm pla}}{\pi G} \, .
\end{equation}
Here  $c$ is the particle velocity dispersion and $\varOmega_{\rm pla}(r)$ is
the Keplerian frequency profile of the circumplanetesimal disc. The velocity
dispersion $c$ is generally not known. Completely elastic particle collisions
cause an exponential transfer of the relative shearing motion of the particles
to random particle motion. The evolution of the random motion stops when the
coefficient of restitution drops at higher collision speeds
\citep{GoldreichTremaine1978}. Estimates of the velocity dispersion in Saturn's
rings motivates setting this transition in the elasticity of the collisions at
$0.001\,{\rm m\,s^{-1}}$. However, the fluffy silicate pebbles that form in
the inner regions of protoplanetary discs may have very
different energy dissipation in inelastic collisions than the ice particles of
Saturn's rings and hence we keep $c$ as a free parameter in the
circumplanetesimal disc model. The profiles of particle column density, particle
scale-height and mid-plane particle density, all calculated for $Q=1$, come out
as
\begin{eqnarray}
  \varSigma_{\rm p} &=& 3.2\,{\rm kg\,m^{-2}}\, \left( \frac{r}{100\,R_{\rm
  pla}} \right)^{-3/2} \left( \frac{c}{0.1\,{\rm m\,s^{-1}}} \right) \left(
  \frac{R_{\rm pla}}{10^2\,{\rm km}} \right)^{3/2} \, , \label{eq:SPd} \\
  H_{\rm p} &=&
  130\,{\rm km}\,\left( \frac{r}{100\,R_{\rm pla}} \right)^{3/2} \left(
  \frac{c}{0.1\,{\rm m\,s^{-1}}} \right) \, , \\
  \rho_{\rm p} &=& 1.1 \times 10^{-3}\,{\rm kg\,m^{-3}} \left(
  \frac{r}{100\,R_{\rm pla}} \right)^{-3}
  \left( \frac{R_{\rm pla}}{10^3\,{\rm km}} \right)^{3/2} \, .
\end{eqnarray}
In these expressions we have normalised the radial distance from the
planetesimal, $r$, by 100 times the planetesimal radius, $R_{\rm pla}$. We plot
the radial profiles of pebble column density, ratio of disc mass to planetesimal
mass, pebble scale-height and pebble mid-plane density for a planetesimal of
radius 100 km and a protoplanet of radius 1000 km in Figure
\ref{f:planetesimal_lightning}. The conditions for the nuclear battery depend
strongly on the adopted value of the random pebble speed $c$ as well as on the
planetesimal size. Discs with random pebble speed in the approximate range
between 0.01 m/s to 0.1 m/s have column densities close to the stopping column
of positrons between 100 and 1000 planetesimals radii where most of the disc
mass resides.

The ratio of the circumplanetesimal disc mass to the mass of the central
planetesimal determines the maximum production efficiency of chondrules. The top
right panel of Figure \ref{f:planetesimal_lightning} shows that the protoplanet
of radius 1000 km has only a small mass fraction in the circumplanetesimal disc
and hence can only produce a relatively small amount of chondrules relative to
its mass. This may nevertheless not necessarily be in conflict with the high
mass fraction of chondrules in chondrites, if these chondrules were produced in
pebble discs around larger protoplanets. As the outer edge of the disc expands
by angular momentum exchange with the inner disc, the chondrules formed in the
circumplanetesimal disc are recycled into the protoplanetary disc and could
contribute significantly to the mass reservoir from which the smaller chondrite
parent bodies formed, even if the production efficiency relative to the
protoplanet mass is low.
\begin{figure*}
  \begin{center}
    \includegraphics[width=0.8\linewidth]{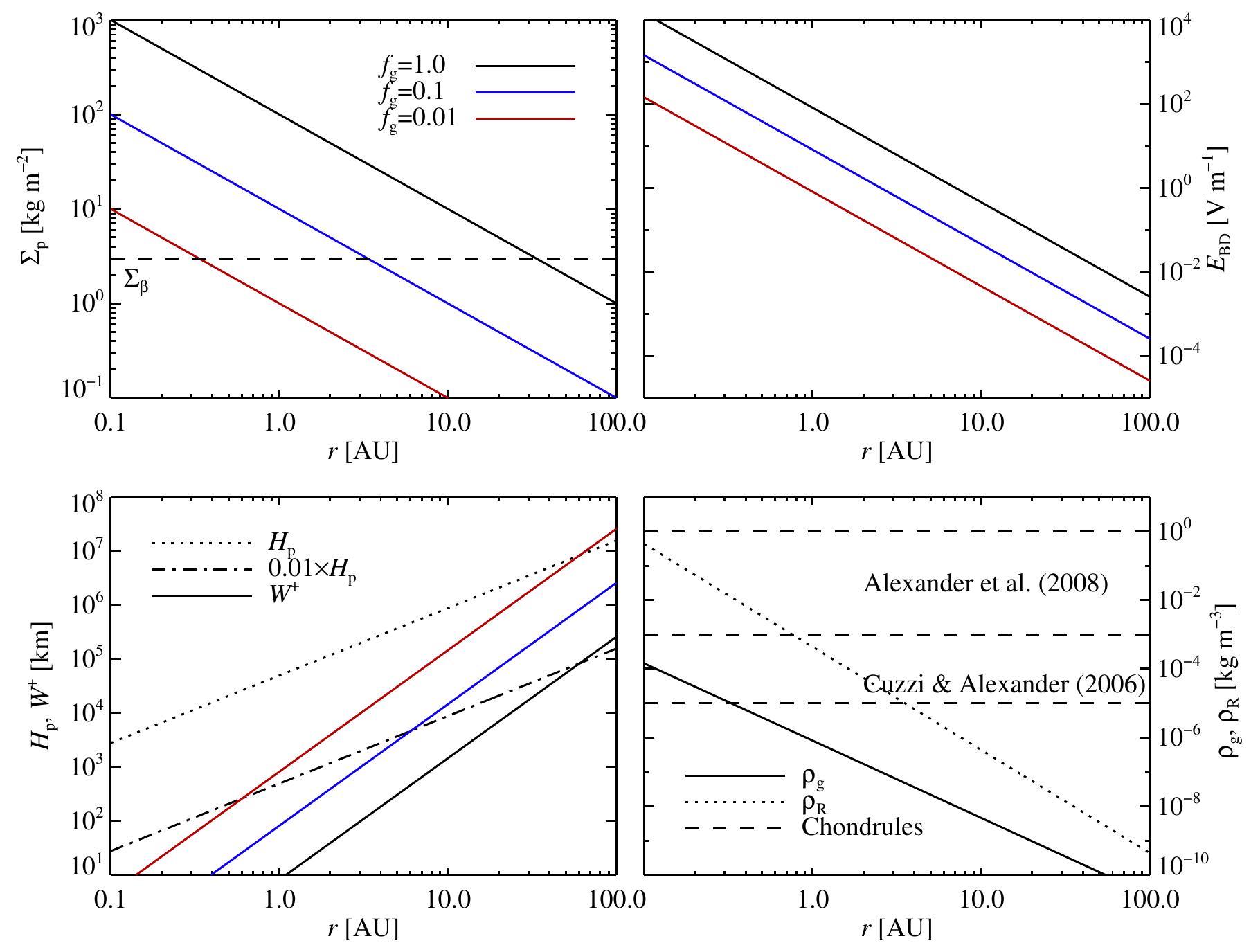}
  \end{center}
  \caption{The conditions for the nuclear battery effect in the protoplanetary
  disc mid-plane. The {\it top left} plot shows the pebble column density for
  three values of gas depletion $f_{\rm g}$ and the stopping column of
  positrons. The mid-plane layer transitions from optically thick to optically
  thin to positron stopping at a gas mass of 10\% of the nominal value in the
  asteroid belt. The {\it top right} plot shows the breakdown value of the
  electric field. The {\it bottom left} plot shows the scale-height of the
  mid-plane layer (dotted line) together with the stopping length of the
  positrons at the breakdown value of the electric field. The mid-plane layer is
  clearly much thicker than the optimal thickness for transferring the positron
  energy to the electric field, but substructures of 1\% of the mid-plane
  thickness are more in agreement with the optimal width (dot-dashed line). The
  {\it bottom right} plot shows the mid-plane density of gas, the Roche density
  and the conditions for chondrule formation. The lack of isotope fractionation
  and stabilisation of melted silicate can be achieved at solids-to-gas ratios
  of $10$ (at 1 AU) to $100$ (at 2.5 AU), but the stricter density conditions
  imposed by the observed high abundance of Na \citep{Alexander+etal2008} can
  only be achieved interior of 1 AU where the Roche density is high.}
  \label{f:disc_lightning}
\end{figure*}

Importantly, the bottom right panel of Figure \ref{f:planetesimal_lightning}
illustrates how the pebble density inside of the Hill radius of the planetesimal
reaches several orders of magnitude above the Roche density in the
protoplanetary disc. Thus the mid-plane conditions around both the planetesimal
and the protoplanet are right in the middle of the constraints inferred by
\cite{Alexander+etal2008} and \cite{AlexanderEbel2012}. The circumplanetesimal
discs that formed in the simulations by \cite{JohansenLacerda2010} had sizes
around 10\% of the Hill radius, or 100 times the planetesimal radius. These
discs would nevertheless not form around small planetesimals, due to the
headwind of the gas. Therefore it is doubtful whether small planetesimals can
host extensive circumplanetesimal discs, unless the gravitational collapse in
itself leads to the formation of a compact, rotationally supported pebble clump
that subsequently expands by viscous stresses into its Hill radius. Such a dense
disc expanding outwards could have enough density to withstand the friction from
the sub-Keplerian gas.  This model is nevertheless very speculative and this
illustrates well the need for computer simulations to understand the formation
and dynamics of circumplanetesimal discs better.

Pebbles in the circumplanetesimal disc will spend most of their time in the
outer regions of the disc where the accretion time-scale is longest. The viscous
time-scale of a circumplanetesimal particle disc of optical depth $\tau$
(=$\varSigma_{\rm p} \sigma_{\rm p}/m_{\rm p}$) is \citep[see
e.g.][]{Charnoz+etal2009}
\begin{equation}
  T = \left( \frac{r}{c} \right)^2 \varOmega_{\rm pla} \frac{1+\tau^2}{\tau}
\end{equation}
The viscous time-scale in the limit $\tau \gg 1$ can now be expressed as
\begin{eqnarray}
  T &=& \left( \frac{r}{c} \right)^2 \varOmega_{\rm pla} \frac{c \varOmega_{\rm
  pla}}{\pi G} \frac{\pi R_{\rm p}^2}{m_{\rm p}} = \frac{R_{\rm pla}^3}{c r
  R_{\rm p}} \nonumber \\ 
  &=& 3.2 \,{\rm Myr}\, \left( \frac{r}{100\,R_{\rm pla}} \right)^{-1} \left(
  \frac{c}{0.1\,{\rm m\,s^{-1}}} \right)^{-1} \left( \frac{R_{\rm
  pla}}{10^3\,{\rm km}} \right)^2 \left( \frac{R_{\rm p}}{1\,{\rm mm}}
  \right)^{-1} \, .
\end{eqnarray}
This is clearly a very long time-scale for a protoplanet of radius $R_{\rm
p}=10^3\,{\rm km}$, which validates our argument that accretion will happen by
gravitational stresses and thus will be regulated by the continuous infall of
new pebble material onto the circumplanetesimal disc. A planetesimal of radius
$R_{\rm p}=100\,{\rm km}$ has a much shorter accretion time-scale, unless the
random pebble motion is less than $0.1\,{\rm m\,s^{-1}}$.

\cite{Johansen+etal2015} showed that planetesimals can grow to protoplanets of
Mars mass, initially by planetesimal-planetesimal collisions and later mainly by
pebble accretion, in the asteroid belt after a few million years. The pebble
accretion process would likely endow these protoplanets with extensive
circumplanetesimal pebble discs. \cite{Drazkowska+etal2016} demonstrated that
the pile up of particles in the terrestrial planet region can trigger
planetesimal formation already after a few hundred thousand years evolution of
the protoplanetary disc. Therefore protoplanets may have populated the
terrestrial planet region even earlier than the asteroid belt region and hence
chondrules could have been produced around these protoplanets and subsequently
transported to the asteroid formation region to become incorporated into the
asteroids and protoplanets that formed there.

\subsection{Nuclear battery in the protoplanetary disc}

The circumplanetesimal disc model presented in the previous subsection is
appealing because the mid-plane pebble density reaches orders of magnitude
higher than the Roche density, as required for heating chondrule precursors to
the melting temperature and avoiding isotopic fractionation and maintaining the
high abundance of Na. This model nevertheless implies that the chondrules found
in chondrites were first processed in the environment near another planetesimal
-- and hence we must evoke a generation of planetesimals and protoplanets that
formed prior to the asteroids in the asteroid belt and produced extensive
amounts of chondrules that became spread throughout the protoplanetary disc.

Here we explore an alternative setting for lightning discharge and thermal
processing of solids, namely the sedimented mid-plane layer in the
protoplanetary disc. This mid-plane layer can not support particle densities in
excess of the Roche density for long time-scales and thus the conditions are at
odds with the formation of chondrules in the asteroid belt (we show below that
the required densities can be reached inside of 1 AU). The mid-plane layer is
nevertheless prone to charging by positron emission and the photons from the
resulting lightning discharge could flash-heat and sublimate material in its
path. These silicate vapours could in turn recondense as tiny matrix particles.
Some authors have actually suggested that mineral growth in chondrules implies
free propagation of the cooling radiation into a region of relatively low pebble
density \citep{MiuraYamamoto2014}. Such conditions would be in agreement with
the mid-plane layer of the protoplanetary disc as a chondrule formation site.

At the nominal ionisation efficiency of $^{26}$Al, we require a mass loading in
small dust, $Z_1$, of at least 50 in the mid-plane layer, to suppress the
neutralising current. The pebble mid-plane layer may nevertheless experience
reduced positron ionisation, since the positrons streaming out from the
mid-plane layer are not replaced by positrons streaming in from below and above
the mid-plane layer, since those regions are underdense in pebbles. Hence the
conductivity will be reduced significantly compared to thick structures. Thus
even a dust-to-gas ratio as low as 10 in the mid-plane layer may yield low
enough conductivities to drive charging of the mid-plane layer faster than the
neutralisation.

We show in Figure \ref{f:disc_lightning} the conditions for the nuclear battery
effect in the sedimented mid-plane layer of pebbles in the protoplanetary disc.
We consider a protoplanetary disc model with gas density $\rho_{\rm g} = f_{\rm
g} \times 10^{-7}\,{\rm kg\,m^{-3}}$ at 2.5 AU, where $f_{\rm g}$ measures the
depletion of gas due to accretion, and a column density profile that falls with
distance from the star $r$ as $\varSigma_{\rm g} \propto r^{-1}$. The mid-plane
layer scale-height is set to $H_{\rm p}=0.01 H$, a typical value when stirred by
the streaming instability \citep{Carrera+etal2015}. We consider three values of
$f_{\rm g}$ -- 1, 0.1 and 0.01 -- to parameterise the time evolution of the
protoplanetary disc.

The column density of pebbles is thick to positron stopping in the interior of
the disc and thin in the exterior. The transition point is at approximately 2.5
AU for a gas density that is 10\% of the nominal value of $10^{-7}\,{\rm
kg\,m^{-3}}$ at 2.5 AU. A mid-plane layer thin to positron stopping is ideal for
the nuclear battery effect, as the positrons from the entire mid-plane layer,
irrespective of any substructure, will be absorbed by the gas outside of the
mid-plane. The mid-plane layer is nevertheless much wider, by two orders of
magnitude at 2.5 AU, than the optimal width for loading positron energy into the
electric field. Therefore substructure of width 1\% of the mid-plane layer width
must be present to reach the breakdown value of the electric field before the
positrons are stopped by the electric field. This implies that the mid-plane
first charges up by sending positrons to the gas above and below the mid-plane
layer, but that the subsequent trapping of positrons in the mid-plane layer
charges up smaller and smaller structures until the breakdown value of the
electric field is reached. Simulations of the streaming instability show that
filaments form on many scales within the mid-plane layer
\citep{JohansenYoudin2007,BaiStone2010a,Johansen+etal2015}. It remains to be
seen whether this density spectrum is prone to charging to the breakdown value
of the electric field; we plan to present such calculations in a future
publication.

The Roche density in the protoplanetary disc is higher inside of 3 AU than the
constraint proposed by \cite{CuzziAlexander2006} to suppress isotopic
fractionation in the heating and stabilise the molten chondrules against
sublimation. The stricter conditions proposed by \cite{Alexander+etal2008} to
main the high abundance of Na within the chondrules can only be achieved below
the Roche density well inside of 1 AU in the disc.

\section{Heating by lightning discharge in a dense pebble region}
\label{s:heating}

We consider in this section the heating effect of lightning discharge into a
dense region of pebbles and dust grains. The solid particles act like a thick,
insulating blanket that facilitates high temperatures and long cooling times for
the relatively moderate discharge powers expected in protoplanetary disc
conditions.  The discharge is assumed to have a total length $h$ into a
cylindrical pebble filament of width $R_*$. The cylindrical symmetry simplifies
the radiative transfer equations significantly. Radiative transfer within a
disc-like geometry is more complex, but we expect that heating by lightning
penetration into a circumplanetesimal disc will be similar to the cylindrical
geometry, as long as the region heated by lightning maintains approximately
cylindrical shape.

\subsection{Radiative equilibrium}

We develop the analytical and numerical heating model around a discrete
radiative-transfer approach where the cylinder is divided into annuli of width
$\Delta R$ corresponding to an optical depth interval of unity. The optical
depth with respect to radiation is dominated by the small dust component. The
mean free path of photons, with wavelengths comparable to or smaller than the
dust so that we can consider geometric absorption, is
\begin{eqnarray}
  \Delta R &=& \frac{1}{n_1 \sigma_1} = \frac{(4/3) R_1 \rho_\bullet}{\rho_1}
  \nonumber \\ &\approx& 1.3\,{\rm m}\,\left( \frac{R_1}{\rm \mu m}\right)
  \left( \frac{\rho_\bullet}{10^3\,{\rm kg\,m^{-3}}} \right) \left(
  \frac{\rho_1}{\rm 10^{-3}\,{\rm kg\,m^{-3}}}\right)^{-1}\, .
\end{eqnarray}
The outer edge of the $i$th cylindrical annulus is denoted $R_i = i \Delta R$.
We assume that each cylindrical ring radiates like a black body whose radiation
is absorbed entirely in its two neighbouring annuli. That gives the temperature
evolution equation for ring $i$ as
\begin{eqnarray}
  \rho_{\rm p} c_{\rm p} V_i \dot{T}_i &=& -2 \pi (R_{i-1} + R_i) h \sigma_{\rm
  SB} T_i^4 + 2 \pi R_{i-1} h \sigma_{\rm SB} T_{i-1}^4 \nonumber \\ & & + 2 \pi
  R_i h \sigma_{\rm SB} T_{i+1}^4 \, .
  \label{eq:Tdot}
\end{eqnarray}
Here $\rho_{\rm p}=\rho_1+\rho_2$ is the total density of solid particles, $c_{\rm p}$ is
the specific heat capacity of the particles, $V_i$ is the volume of annulus $i$,
$\sigma_{\rm SB}$ is the Stefan-Boltzmann constant and $T_i$ is the temperature
of annulus $i$. We then introduce the source function $S_i = \sigma_{\rm SB}
T_i^4$.
The energy flux over the edge of annulus $i$ is
\begin{equation}
  \mathcal{F}_i = 2 \pi R_i h S_i - 2 \pi R_i h S_{i+1} \, .
\end{equation}
In equilibrium this equals the energy flux release by the lightning,
$\mathcal{F}_i = I E h$. Here $I$ is the current carried by the lightning and
$E$ the breakdown electric field. The continuous equation for the energy flux is
\begin{equation}
  \mathcal{F} = -2 \pi R h \Delta R \frac{\dpa S}{\dpa R} = I E h \, .
  \label{eq:Frad}
\end{equation}
We define the optical depth $\tau$ measured inwards from the edge of the pebble
region, with a total optical depth of $\tau_\star$. The optical depth thus
relates to the radial coordinate $R$ through $R = \Delta R (\tau_\star-\tau)$.
Rewriting equation (\ref{eq:Frad}) in terms of optical depth $\tau$ we get
\begin{equation}
  \mathcal{F} = 2 \pi h \Delta R (\tau_\star-\tau) \Delta R \frac{\dpa S}{\Delta
  R \dpa \tau} = I E h \, .
\end{equation} 
Rearranging we find
\begin{equation}
  (\tau_\star-\tau) \frac{\dpa S}{\dpa \tau} = \frac{I E}{2 \pi \Delta R} \, .
\end{equation}
Integration yields
\begin{equation}
  S(\tau) = C - \frac{I E}{2 \pi \Delta R} \ln(\tau_\star-\tau)
\end{equation}
The constant $C$ is fixed by requiring that all the released energy is radiated
outwards at the $\tau=0$ surface, corresponding to the requirement $ \tau_\star
S(1/2) = I E / (2 \pi \Delta R)$ in the outermost annulus with $\tau=1/2$.  This
gives the final source function expression
\begin{equation}
  S(\tau) =  \frac{I E}{2 \pi \Delta R} \left[ \frac{1}{\tau_\star} +
  \ln(\tau_\star-1/2)-\ln(\tau_\star-\tau) \right] \, .
  \label{eq:Stau}
\end{equation}
The temperature at optical depth $\tau=\tau_\star-1/2$, the approximate optical
depth of the first annulus, in the limit $\tau_\star \gg 1$, is
\begin{eqnarray}
  T_\star &=& 1754\,{\rm K} \, \left( \frac{I}{10^5\,{\rm A}} \right)^{1/4}
  \left( \frac{\rho_{\rm g}}{10^{-7}\,{\rm kg\,m^{-3}}} \right)^{1/4} \left(
  \frac{\rho_\bullet}{10^3\,{\rm kg\,m^{-3}}} \right)^{-1/4} \nonumber \\ & &
  \times \left( \frac{R_1}{10^{-6}\,{\rm m}} \right)^{-1/4} \left(
  \frac{\rho_1}{10^{-3}\,{\rm kg\,m^{-3}}} \right)^{1/4} \left(
  \frac{\ln[2 \tau_\star]}{\ln[200]} \right)^{1/4} \, .
  \label{eq:Tstar}
\end{eqnarray}
We used here a typical current carried by terrestrial lightning and landed
fortuitously in the ballpark temperature for chondrule heating. The
characteristic density of protoplanetary discs is many orders of magnitude lower
than the density of the terrestrial atmosphere. The discharge process may
nevertheless be taken to follow the so-called Townsend scalings
\citep{Ebert+etal2010}, which yield a discharge length-scale proportional to the
number density of neutrals $n$, a discharge time-scale proportional to $1/n$ and
a discharge current independent of $n$. \cite{HoodKring1996} used the 100 ms
duration of terrestrial lightning to argue that protoplanetary disc lightning
has a characteristic time-scale of $10^4$--$10^5$ s. However, terrestrial
lightning strokes consist of several flashes of $30$ $\mu$s duration, which
would yield $10^2$ s time-scales in the asteroid belt region of the
hydrogen-dominated solar protoplanetary discs with $\rho_{\rm g}=10^{-7}\,{\rm
kg\,m^{-3}}$.

Note that the central temperature in equation (\ref{eq:Tstar}) has a very weak
scaling with $\tau_\star$ -- even a dramatic increase in the optical depth will
not lead to a significant temperature increase in the core of the filament. This
is a result of the radial thinning of the radiation at large distances from the
lightning. The temperature at the optical at the optical surface ($\tau=1/2$)
relates to the deep temperature $T_\star$ through
\begin{equation}
  T(1/2) = \frac{T_\star}{[\tau_\star \ln (2 \tau_\star)]^{1/4}} \, .
\end{equation}
That gives $T(1/2)=366$ K for $T_\star=1754$ K and $\tau_\star=100$. The
total energy content of the pebble region at the equilibrium temperature is
obtained by integrating over the energy contents of the cylindrical annuli,
\begin{equation}
  Q = \int_0^{R_\star} c_{\rm p} \rho_{\rm p} T 2 \pi R h \de R =
      2 \pi h c_{\rm p} \rho_{\rm p} (\Delta R)^2 \int_{0}^{\tau_*} T \tau \de
      \tau \, .
\end{equation}
Inserting equation (\ref{eq:Stau}) we obtain the expression
\begin{eqnarray}
  Q &=& \left( \frac{I E}{2 \pi \sigma_{\rm SB} \Delta R} \right)^{1/4} 2 \pi h
  c_{\rm p} \rho_{\rm p} (\Delta R)^2 \nonumber
  \\ & & \times \int_0^{\tau_*} \left[ \frac{1}{\tau_\star} + \ln
  (\tau_\star-1/2) - \ln(\tau_\star-\tau) \right]^{1/4} \tau \de \tau \, .
\end{eqnarray}
We find empirically that the term in hard brackets within the integral has a
mean value, weighted by $\tau$, of approximately unity for a wide range of
values of $\tau_\star$. Hence the integral is well approximated by $(1/2)
\tau_\star^2$. That gives the energy content per length
\begin{eqnarray}
  Q/h &=& 6.5 \times 10^8\,{\rm J\,m^{-1}} \times \left( \frac{\tau_\star}{100}
  \right)^2 \nonumber \\ & & \times \left(
  \frac{I}{10^5\,{\rm A}} \right)^{1/4} \left( \frac{\rho_{\rm g}}{10^{-7}\,{\rm
  kg\,m^{-3}}} \right)^{1/4} \nonumber \\
  & & \times \left( \frac{c_{\rm p}}{10^3\,{\rm J\,kg^{-1}\,K^{-1}}} \right)
  \left( \frac{\rho_2}{10^{-2}\,{\rm kg\,m^{-3}}} \right)
  \left( \frac{\rho_1}{10^{-3}\,{\rm kg\,m^{-3}}} \right)^{-7/4} \nonumber
  \\ & & \times \left( \frac{R_1}{10^{-6}\,{\rm m}} \right)^{7/4} \left(
  \frac{\rho_\bullet}{10^{3}\,{\rm kg\,m^{{-3}}}} \right)^{7/4} \, .
\end{eqnarray}
Here we assumed that the density of solid particles is dominated by the pebble
component, $\rho_{\rm p} \approx \rho_2$. The time to heat the dense pebble
region is
\begin{eqnarray}
  t_{\rm heat} &=& \frac{Q/h}{I E} = 7.7 \times 10^2\,{\rm s}\, \left(
  \frac{\tau_\star}{100} \right)^2 \nonumber \\
  & & \times \left(
  \frac{I}{10^5\,{\rm A}} \right)^{-3/4} \left( \frac{\rho_{\rm g}}{10^{-7}\,{\rm
  kg\,m^{-3}}} \right)^{-3/4} \nonumber \\
  & & \times \left( \frac{c_{\rm p}}{10^3\,{\rm J\,kg^{-1}\,K^{-1}}} \right)
  \left( \frac{\rho_2}{10^{-2}\,{\rm kg\,m^{-3}}} \right)
  \left( \frac{\rho_1}{10^{-3}\,{\rm kg\,m^{-3}}} \right)^{-7/4} \nonumber
  \\
  & & \times 
  \left( \frac{R_1}{10^{-6}\,{\rm m}} \right)^{7/4}
  \left( \frac{\rho_\bullet}{10^{3}\,{\rm kg\,m^{{-3}}}} \right)^{7/4} \, .
  \label{eq:theat}
\end{eqnarray}

\subsection{Heating and cooling rates}

\begin{figure}
  \begin{center}
    \includegraphics[width=\linewidth]{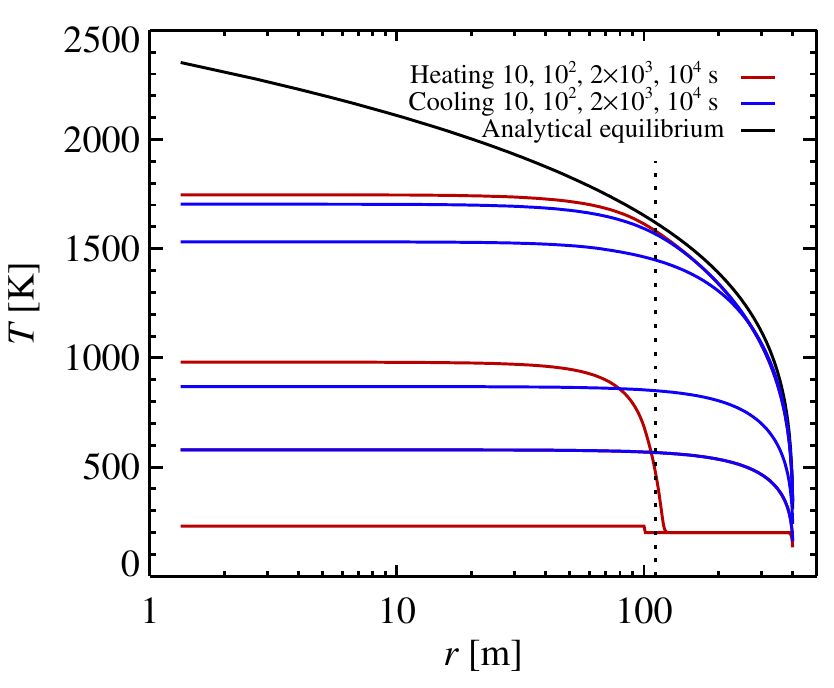}
  \end{center}
  \caption{The temperature as a function of distance from the centre of the
  lightning discharge channel for a current $I=10^5\,{\rm A}$, gas density
  $\rho_{\rm g} = 3 \times 10^{-7}\,{\rm kg\,m^{-3}}$, dust density
  $\rho_1=10^{-3}\,{\rm kg\,m^{-3}}$, pebble density $\rho_2=10^{-2}\,{\rm
  kg\,m^{-3}}$ and optical depth $\tau_\star=300$. The discharge channel has a
  width of 100 m, indicated by the vertical dotted line. The discharge is
  assumed to last $2 \times 10^3$ s, followed by a cooling phase that is evolved
  up to a time of $10^4$ s. Heating is rapid and uniform within the discharge
  channel, while the outer regions only heat after radiation diffuses out from
  the discharge region. The equilibrium temperature follows the analytical
  solution outside of the discharge channel.}
  \label{f:lightning_temperature_vs_time}
\end{figure}
\begin{figure}
  \begin{center}
    \includegraphics[width=\linewidth]{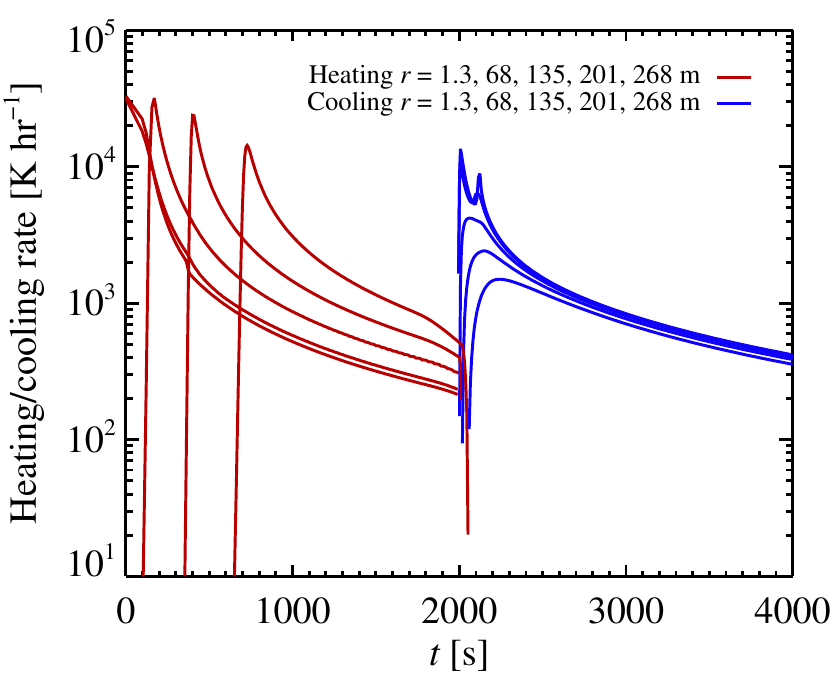}
  \end{center}
  \caption{Heating rate (red) and cooling rate (blue) of the pebble filament
  heated by the lightning discharge shown in Figure
  \ref{f:lightning_temperature_vs_time}. The curves show, from top to bottom,
  the heating and cooling rate at optical depth $\tau=0,50,100,150,200$. The
  cooing rates lie at a few times $10^3\,{\rm K\,hr^{-1}}$ in the first 100
  seconds after heating terminates at $t_{\rm heat}=2\times10^{3}\,{\rm s}$ and
  subsequently fall to below $10^3\,{\rm K\,hr^{-1}}$ over the next 1000 s.
  These cooling rates are in good agreement with the upper range of cooling
  rates inferred from chondrule textures \citep{HewinsRadomsky1990}.}
  \label{f:lightning_cooling_rate_vs_time}
\end{figure}
Chondrule precursors appear to have been heated to near-liquidus temperatures,
reaching the range 1500 K -- 2200 K \citep{HewinsRadomsky1990}. This constraint
is mainly based on what appears to be relict, unmelted grains inside chondrules.
If these grains are instead interpreted as interlopers that penetrated a fully
molten spherule \citep{ConnollyHewins1995}, then chondrule precursors could have
been heated to much higher temperatures. The cooling rates of chondrules can be
constrained by studying crystal growth in furnace experiments of chondrule
analogues or from the retention of volatiles such as Na in chondrules.
\cite{HewinsRadomsky1990} found that cooling rates in the range $10^2$--$10^3$
K/hr are consistent with experimental data. The upper range of these cooling
rates corresponds to what is inferred for matrix grains \citep{ScottKrot2005}.
On the other hand, \cite{YurimotoWasson2002} and \cite{MiuraYamamoto2014}
inferred much higher rates, in the range $10^5$--$10^7$ K/hr. The upper range of
those rates are comparable to cooling rate of isolated molten droplets,
\begin{eqnarray}
  \dot{T} &=& \frac{4 \pi R_2^2 \sigma_{\rm SB} T^4}{c_{\rm p} (4 \pi/3)
  \rho_\bullet R_2^3} \nonumber \\ &\approx& 9.8 \times 10^6\,{\rm
  K\,s^{-1}}\,\left( \frac{T}{2000\,{\rm K}} \right)^4 \left(
  \frac{\rho_\bullet}{10^3\,{\rm kg\,m^{-3}}} \right)^{-1} \left(
  \frac{R_2}{1\,{\rm mm}} \right)^{-1} \, .
\end{eqnarray}
Much lower cooling rates can be achieved in an optically thick region of
chondrules. We measure the heating and cooling rates of pebbles exposed to
radiation from lightning discharge by solving equation (\ref{eq:Tdot})
numerically starting at a background temperature of $T=200$ K. Absorption
and release of latent heat $\mathcal{L}$ by melting and crystallisation is
included using a simplified approach where we translate the internal energy $e$
to temperature $T$ through the relations
\begin{eqnarray}
  T &=& \frac{e}{\rho_2 c_{\rm p}} \quad {\rm for} \quad e \le e_{\rm s} \, , \\
  T &=& T_{\rm s} + \frac{T_{\rm l}-T_{\rm s}}{\rho_2 [\mathcal{L} + c_{\rm p}
  (T_{\rm l}-T_{\rm s})]} (e - e_{\rm s}) \quad {\rm for} \quad e_{\rm s} < e
  \le e_{\rm l} \, , \\
  T &=& \frac{(e-\rho_2 \mathcal{L})}{\rho_2 c_{\rm p}} \quad {\rm for} \quad e >
  e_{\rm l} \, .
\end{eqnarray}
Here $T_{\rm s}$ is the solidus temperature (taken to be 1500 K), $T_{\rm l}$ is
the liquidus temperature (taken to be 2000 K), and the internal energies at
solidus and liquidus, respectively, are defined as
\begin{eqnarray}
  e_{\rm s} &=& \rho_2 c_{\rm p} T_{\rm s} \, , \\
  e_{\rm l} &=& \rho_2 c_{\rm p} T_{\rm l} + \rho_2 \mathcal{L} \, .
\end{eqnarray}
Absorption and release of latent heat does not change the equilibrium
temperature profile, as given in equation (\ref{eq:Stau}), but delays both the
initial heating and the subsequent cooling.

The width of the lightning discharge channel must be scaled properly to the
protoplanetary disc conditions. \cite{HoranyiRobertson1996} argued that the
discharge channel has a width of 1,000-10,000 times the mean free path of the
electrons. This scaling relation yields
\begin{equation}
  W_{\rm dis} = 334\,{\rm m}\, \left( \frac{f_{\rm dis}}{10^3} \right) \left(
  \frac{\rho_{\rm g}}{10^{-7}\,{\rm kg\,m^{-3}}} \right)^{-1}
\end{equation}
Here $f_{\rm dis}$ denotes the discharge width in units of the mean free path of
electrons. We use here $f_{\rm dis}=10^3$ and note that the equilibrium
temperature solution of equation (\ref{eq:Stau}) is still valid outside the
discharge channel, while the temperature within the discharge channel is lower
than the equilibrium solution as the decreased heat flux necessitates a lower
temperature gradient.

In Figure \ref{f:lightning_temperature_vs_time} we show the evolution of the
temperature profile of the pebble filament for selected times. We assume that
the lightning discharge heats for a time-scale of $t_{\rm
heat}=2\times10^3\,{\rm s}$.  The temperature climbs steadily during this time,
but only approaches the analytical equilibrium solution (equation \ref{eq:Stau})
outside the discharge channel. The temperature is relatively constant within the
discharge channel itself and significantly below the analytical solution due to
the lower energy flux.  Cooling occurs relatively isothermally over the pebble
filament. In Figure \ref{f:lightning_cooling_rate_vs_time} we show the heating
and cooling rate at selected regions in the pebble filament as a function of
time. Heating is initially rapid, but slows down as the equilibrium temperature
is approached.  The cooling rate lies at a few times $10^3\,{\rm K\,hr^{-1}}$ in
the first few hundred seconds after heating terminates and subsequently falls to
a few times $10^2\,{\rm K\,hr^{-1}}$ as the temperature drops to 1000 K. These
cooling times lie towards the upper end of the values inferred from chondrule
textures \citep{HewinsRadomsky1990}. The cooling time of the filament should
broadly follow the heating time, so following equation (\ref{eq:theat}) lower
cooling rates could be achieved in filaments of large optical depth or lower gas
density.

\section{Summary}
\label{s:summary}

We have demonstrated in this paper that the decay energy of the short-lived
radionuclide $^{26}$Al can be harvested to drive lightning discharge in
protoplanetary discs. Positrons released from dense pebble structures load their
energy into the electric field that builds up by the negative charging of these
structures. This process is akin to a direct-charging nuclear battery where a
radioactive isotope charges a capacitor. The total energy available in the
kinetic energy of the released positrons is comparable to the energy required to
melt all the solid material in a protoplanetary disc. Such an immense energy
released in lightning discharge may be observable in nearby protoplanetary discs
with e.g.\ the ALMA telescopes \citep{Muranushi+etal2015}.

The neutralising gas current, driven with an efficiency (conductivity) that is
primarily set by positron ionisation of hydrogen molecules, can no longer
compete with the positron current when the dust-to-gas ratio of small grains is
higher than a few tens. This leaves lightning discharge as the natural path to
charge neutralisation, releasing the energy stored in the electric field in thin
channels. \cite{Eisenhour+etal1994} demonstrated that heating by infrared laser
light yields melting primarily of pebbles larger than 0.1 mm, due to the
inability of smaller pebble aggregates in absorbing infrared photons. In this
picture, the largest chondrules represent the largest available chondrule
precursors pebbles, while the lack of small chondrules in most chondrites
reflects inefficient heating of small pebbles by the heat front propagating from
the lightning channel.

We find that the optimal conditions for heating pebbles by lightning discharge
arise in dense particle discs that orbit around planetesimals and protoplanets.
The high dust content of such discs implies a low conductivity and the rapid
transition from dense pebble environment to the surrounding protoplanetary disc
dominated by gas is ideal for driving an outwards directed current of positrons.
Circumplanetesimal discs are appealing from a cosmochemical point of view also,
since the prevalent conditions there match the high dust densities required for
chondrule melting and high abundance of volatiles
\citep{CuzziAlexander2006,Alexander+etal2008}. The formation and dynamics of
such circumplanetesimal discs is nevertheless not well understood and should be
the focus of future research.

The sedimented mid-plane layer of pebbles in the protoplanetary disc can also
become charged by the emission of positrons, if the mass loading in small dust
grains is high enough in the mid-plane (larger than 10--50) to reduce the
neutralising current. The densities of solids in the mid-plane layer are too low
in the region of the asteroid belt to allow heating to the liquid phase and the
solids are expected to sublimate completely upon flash heating. The subsequent
cooling and recondensation could form small particles similar to the matrix
grains found between chondrules in chondrite meteorites, although the long
cooling times inferred for some matrix grains may imply that matrix and
chondrules formed under elevated densities \citep{ScottKrot2005}. The density in
the mid-plane is much higher in sub-AU distances than in the asteroid belt, so
the conditions close to the star are more in agreement with the cosmochemical
constraints on the elevated density of chondrule precursors

The solar protoplanetary disc appears to have been injected with a nominal value
of $^{26}$Al in its birth cluster. Stellar systems are born with a range of
abundances from ten times below to ten times above the solar abundance
\citep{Vasileiadis2013,Parker+etal2014,Lichtenberg+etal2016}. Our results show
that the energy budget for chondrule melting could have been supplied by the
decay of $^{26}$Al. The pebbles in the rare systems with ten times the solar
abundance of $^{26}$Al must have undergone widespread and repeated melting that
could even affect the water contents, and hence the habitability, of terrestrial
planets, if ices are continuously driven off the pebbles in large regions of the
protoplanetary disc.  Therefore it is imperative to understand the heating
mechanism that processes solids in protoplanetary discs to form chondrules and
matrix. We have demonstrated here that the decay of $^{26}$Al -- already
generally accepted to be the energy source for melting and differentiating
asteroids -- may also have provided the energy needed for the widespread thermal
processing and melting of dust and pebbles in protoplanetary discs.

\begin{acknowledgements}

AJ thanks the Knut and Alice Wallenberg Foundation (grants 2012.0150, 2014.0017,
2014.0048), the Swedish Research Council (grant 2014-5775) and the European
Research Council (ERC Starting Grant 278675-PEBBLE2PLANET and ERC Consolidator
Grant 724687-PLANETESYS) for their financial support. SO is supported by the
JSPS Grants-in-Aid for Scientific Research (Grant Number JP17K18812).  The
authors are thankful to the referee, Steven Desch, for a thorough referee report
that led to many improvements in the presentation of the paper.

\end{acknowledgements}

\appendix

\section{Charge equations}
\label{s:charge}

In this appendix we formulate the equations for production of electrons and ions
by high-energy positrons and the charging of dust particles. In the following
appendix, Appendix \ref{s:equilibrium}, we will find the equilibrium state for
electron and ion densities and the charge of dust particles and calculate the
conductivity of the gas and the neutralising current as a function of the
electric field and compare this to the current driven by the radial drift of the
pebbles.

\subsection{Ionisation rates}

The evolution equations for the electron and ion densities are
\begin{eqnarray}
  \dot{n}_{\rm e} &=& n_\beta n_{\rm n} \langle \sigma_{\rm n\beta} v_\beta
  \rangle - n_{\rm e} n_d K_{d{\rm e}} 
  + N_{{\rm s}d} n_\beta n_d \sigma_d v_\beta
  \label{eq:nedot} \, ,
  \\ \dot{n}_{\rm i} &=& n_\beta n_{\rm n} \langle \sigma_{\rm n\beta} v_\beta
  \rangle 
  - n_{\rm i} n_d K_{d{\rm i}} + n_\beta n_{\rm n}
  \langle \sigma_{\rm n\beta} v_\beta \rangle (E_{\rm ion} / T_\beta) \label{eq:nidot} \, .
\end{eqnarray}
The first term appearing in both Equations (\ref{eq:nedot}) and
(\ref{eq:nidot}) represents gas ionisation by the passing positron (here
$\sigma_{\rm n\beta}$ and $v_\beta$ are the cross sections of ionisation and
the speed of the positrons). The second term represents dust absorption of
electrons and ions. The index $d$ implies summation over species. The
adsorption rates of electrons and ions onto the dust grains (absorption cross
section of dust times speed of the charged species) are denoted $K_{d{\rm e}}$
and $K_{d{\rm i}}$, respectively. These rates are complicated functions of the
charge of the dust; we derive expressions for $K_{d{\rm e}}$ and $K_{d{\rm i}}$
in Appendix \ref{s:Kdi}.

The third term of Equation (\ref{eq:nedot}) represents the emission of $N_{{\rm
s}d}$ secondary electrons as a positron passes the surface of a particles. In
order to obtain the values of $N_{{\rm s}d}$, averaged over impact parameter, we
have performed a Monte Carlo simulation of the penetration of a positron through
a solid particle of finite radius. The results are discussed in Appendix
\ref{s:secondary}. We find $N_{{\rm s}1}=0.04$ for micrometer-sized particles
and $N_{{\rm s}2}=0.08$ for millimeter-sized pebbles. Secondary electrons leave
silicon targets with typical energies of approximately 10 eV
\citep{Dapor+etal2008}; this energy is matched by the electric potential energy
from a positive charge of $+10^7 e$ on a millimeter-sized pebble. Therefore we
can safely assume for all pebble charges found in Section \ref{s:equilibrium}
that secondary electrons can freely leave the surface of the pebbles.

The third term of Equation (\ref{eq:nidot}) represents the absorption of
positrons by a gas molecule as the positron comes to rest, to create a
$\gamma$-ray when it combines with an electron. The rate of this event equals
the ionisation rate by the positron times the ratio of the ionisation energy of
the hydrogen molecule, taken to be $E_{\rm ion}=37$ eV, to the kinetic energy of
the positron ($T_\beta=0.66$ MeV for the decay of $^{26}$Al).

\subsection{Dust charging}

The charging of small dust grains and macroscopic pebbles follows the evolution
equations
\begin{eqnarray}
  \dot{q}_1 &=& - n_{\rm e} K_{1{\rm e}} + n_{\rm i} K_{1{\rm i}} + N_{\rm s1}
  n_\beta \sigma_1 v_\beta - r_\beta m_1^* \nonumber \\
  & & +\, n_\beta \sigma_1 v_\beta (R_1^*/\chi_{\rm s}) + n_2 q_1 K_{12} \, ,
  \label{eq:q1dot} \\
  \dot{q}_2 &=& - n_{\rm e} K_{2{\rm e}} + n_{\rm i} K_{2{\rm i}} + N_{\rm s2}
  n_\beta \sigma_2 v_\beta - r_\beta m_2^* \nonumber \\
  & & \, + n_\beta \sigma_2 v_\beta (R_2^*/\chi_{\rm s}) + n_1 q_1 K_{12} \, .
  \label{eq:q2dot}
\end{eqnarray}
The first two terms represent collisions with electrons and ions, the third term
the production of secondary electrons, the fourth term the charging by release
of positrons ($r_\beta$ is the positron production rate per unit mass) and the
fifth term is the absorption of positrons ($\chi_{\rm s}\approx 1.3$ mm is the
stopping length of the positrons in the solid particles).

The last term present in both equation (\ref{eq:q1dot}) and (\ref{eq:q2dot})
represents the transfer of charge from small dust grains to pebbles by
collisions \citep{Okuzumi2009}. We adopt here the closure that the number
density of dust and pebbles remains constant, under the assumption that small
dust grains transfer all their charge to a pebble following a bouncing
collision. The adsorption rates $K_{d{\rm e}}$, $K_{d{\rm i}}$ and $K_{12}$ are
described in Appendix \ref{s:Kdi}.

\subsection{Positron density and charge equilibrium}

The positron number density follows the equation
\begin{equation}
  \dot{n}_\beta = r_\beta \rho_{\rm p}^* - n_\beta n_d \sigma_d v_\beta
  (R_d^*/\chi_{\rm s}) - n_\beta n_{\rm n} \langle \sigma_{\rm n} v_\beta
  \rangle (E_{\rm ion}/T_\beta) \, .
  \label{eq:nbetadot}
\end{equation}
The first term is the emission rate, with $\rho_{\rm p}^* = n_1 m_1^* + n_2
m_2^*$. The second term is the absorption in solid particles and the third
term the absorption in gas molecules. All these terms are balanced in
corresponding terms in the evolution of the number densities of electrons and
ions and the charging of dust. We see that the charge is conserved, since
\begin{equation}
  - \dot{n}_{\rm e} + \dot{n}_{\rm i} + \dot{n}_\beta + \dot{q}_1 n_1 +
    \dot{q}_2 n_2 = 0 \, .
    \label{eq:ccons}
\end{equation}
The integral of the equation is $Q$, the free charge density, which we set to be
zero. This means that any of the four dynamical equations can be written as a
sum over the other three, so that the number of independent dynamical equations
is only three.

\begin{figure}
  \begin{center}
    \includegraphics[width=0.9\linewidth]{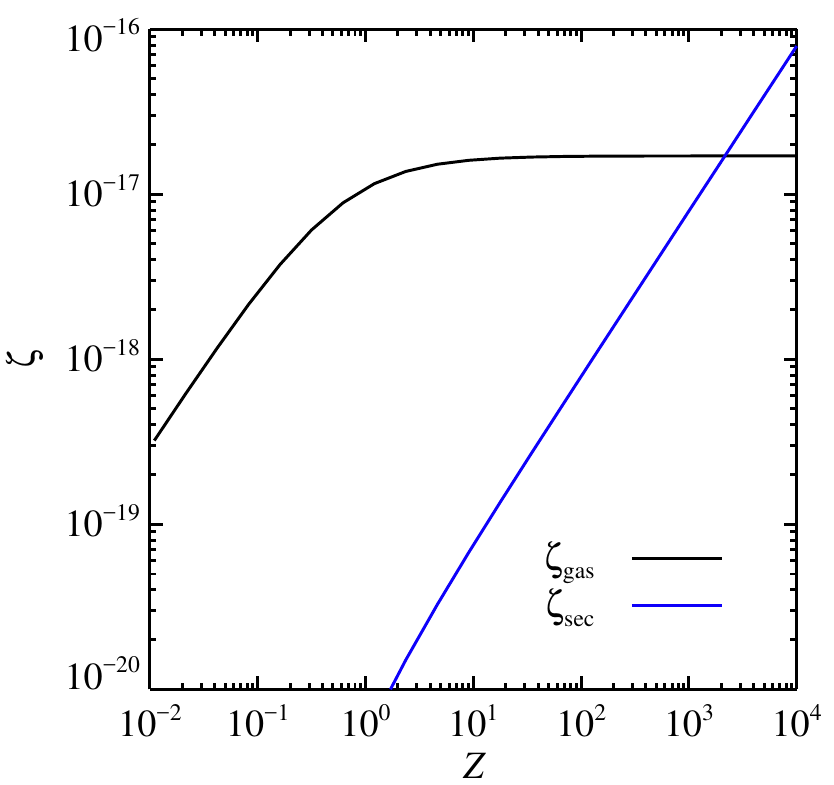}
  \end{center}
  \caption{The ionisation rate $\zeta$ versus the dust-to-gas ratio. The
  initial rise of the ionisation rate is due to the increased density of
  $^{26}$Al, while the plateau occurs when the rise is balanced by an increased
  absorption rate of positrons in dust grains. For the production of secondary
  electrons, which occurs in collisions with dust, we have converted to an
  effective rate on the gas particles for comparison. Secondary electrons come
  to dominate after $Z \approx 1000$.}
  \label{f:zeta_Z}
\end{figure}

The ionisation rate by positrons is equivalent to
\begin{equation}
  \zeta = n_\beta \langle \sigma_{\rm n\beta} v_\beta \rangle \, .
\end{equation}
We plot the effective ionisation rate $\zeta$ of positrons in Figure
\ref{f:zeta_Z} as a function of the mass loading of dust $Z=\rho_{\rm
p}/\rho_{\rm g}$, calculated from the equilibrium state of equation
(\ref{eq:nbetadot}). The ionisation rate increases initially with $Z$, as the
concentration of $^{26}$Al increases relative to the gas. However, above $Z=1$
the ionisation rate flattens out. At those high particle densities the increase
in positron emission rate is balanced by the stopping of the positrons in the
solid particles. We also show in Figure \ref{f:zeta_Z} the production rate of
secondary electrons, calculated as an effective ionisation rate of the gas. The
contribution of secondary electrons to the electron density dominates at mass
loading above $Z \sim 1000$.

\subsection{Electron and ion heating}
\label{s:eiheating}

The speed of the electrons, of mass $m_{\rm e}$, and the ions, of mass $m_{\rm
i}$, can be obtained from their temperatures $T_{\rm e}$ and $T_{\rm i}$ through
\begin{eqnarray}
  \frac{1}{2} m_{\rm e} v_{\rm e}^2 &=& \frac{3}{2} k_{\rm B} T_{\rm e} \, , \\
  \frac{1}{2} m_{\rm i} v_{\rm i}^2 &=& \frac{3}{2} k_{\rm B} T_{\rm i} \, .
\end{eqnarray}
\cite{OkuzumiInutsuka2015} showed that the electrons and ions obtain energies
above the mean thermal energy of the molecules in the presence of strong
electric fields. This is due to transfer of energy from the mean motion along
the electric field to random motion after collisions with neutrals. The
increased random motion in term decreases both the number densities and mean
current speed of electrons and ions. Following \cite{MoriOkuzumi2016} we write
the electron temperature as
\begin{equation}
  T_{\rm e} = T \left[ \frac{1}{2} + \sqrt{\frac{1}{4}+\frac{2}{3} \left(
  \frac{E}{E_{\rm crit}} \right)^2} \right] \, .
\end{equation}
The critical electric field for electron heating $E_{\rm crit}$ is
\begin{equation}
  E_{\rm crit} = \sqrt{\frac{6 m_{\rm e}}{m_{\rm n}}} \frac{k_{\rm B} T}{e
  \ell_{\rm e}} \, .
\end{equation}
Here $m_{\rm e}$ and $m_{\rm n}$ are the electron and neutral masses,
respectively. The mean free path of the electrons is simply $\ell_{\rm e} =
(n_{\rm n} \sigma_{\rm en})^{-1}$ with $\sigma_{\rm en}=10^{-19}$ m$^2$.

The ions are heated at higher values of the electric field.
\cite{MoriOkuzumi2016} give an expression for the ion temperature,
\begin{equation}
  T_{\rm i} = T \left[ 1 + 7.6 \times 10^{-7} \left( \frac{T}{100\,{\rm K}}
  \right) \left( \frac{E}{E_{\rm crit}} \right)^2 \right] \, .
\end{equation}
The ion temperature increases as the square of the electric field, while the
electron temperature only increases linearly with the electric field in the
limit of high field strength.
\begin{figure*}
  \begin{center}
    \includegraphics[width=0.8\linewidth]{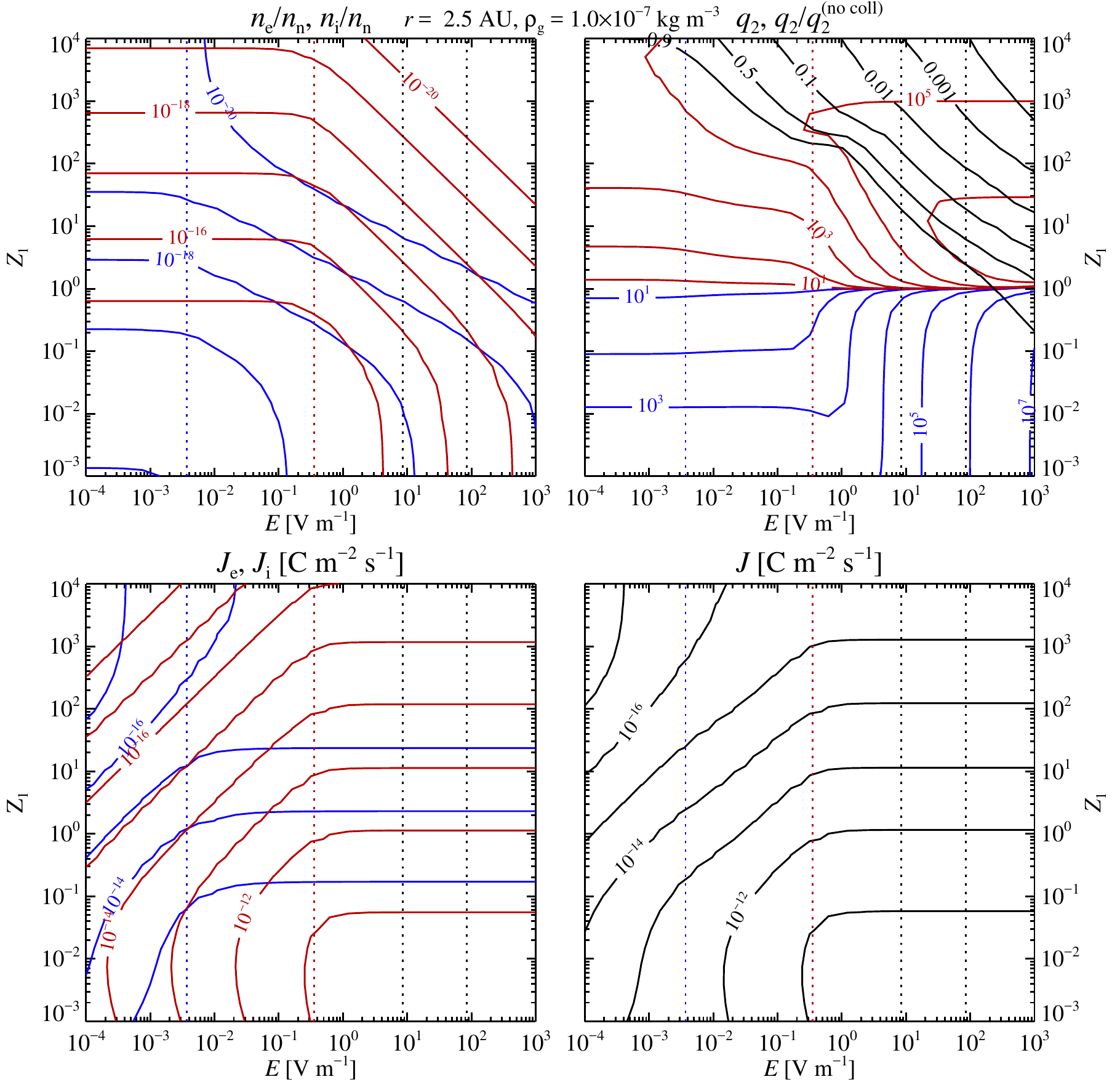}
  \end{center}
  \caption{The electron and ion densities ({\it top left}; blue lines indicate
  electrons and red lines ions), the pebble charge ({\it top right}, blue lines
  indicate negative charge, red lines positive charge, black lines the charge
  relative to a model that neglects charge transfer in dust collisions), the
  electron and ion currents ({\it bottom left}) and the total current ({\it
  bottom right}). The electron and ion critical fields are marked in blue and
  red dotted lines, while the break down electric field is shown with black
  dotted lines (we show both the nominal value and a value ten times lower
  relevant when taking into account the distribution of electron energies). The
  positive pebble charging is mainly due to the emission of secondary electrons
  as positrons penetrate dust surfaces. The equilibrium charge increases with
  increasing field strength, since the field accelerates electrons and ions to
  high speeds, thus increasing the collision rate with the charged pebbles.}
  \label{f:charge}
\end{figure*}

\section{Charge equilibrium}
\label{s:equilibrium}

The direct temporal integration to equilibrium of the charging equations
presented in Appendix \ref{s:charge} is made very expensive by the large range of
time-scales present in the problem. Particularly at high particle mass loading
and strong electric fields, relevant for the heating of chondrules, the
time-scale for collisions between electrons and dust particles is very low.
Therefore we search for the equilibrium condition by solving directly for
$\dot{n}_{\rm e}=\dot{n}_{\rm i}=\dot{q}_1=\dot{q_2}=0$. Unfortunately, none of
these dynamical equations are analytically solvable, due to the complicated
expressions for the adsorption rates $K_{d{\rm e}}$ and $K_{d{\rm i}}$ (see
Appendix \ref{s:Kdi}).

\subsection{Numerical solution method}

One of the four variables in the solution vector $(n_{\rm e},n_{\rm i},q_1,q_2)$
can be eliminated by using the charge conservation condition from equation
(\ref{eq:ccons}). Our algorithm for finding the equilibrium state consists of
stepping through $n_{\rm i}$ from $10^{-30} n_{\rm n}$ to $10^{-5} n_{\rm n}$.
For each value of $n_{\rm i}$ we find $n_{\rm e}$ that satisfies $\dot{n}_{\rm
e}=0$. This requires additional knowledge of $q_1$ and $q_2$. The value of
$q_1$, given $n_{\rm e}$ and $n_{\rm i}$, is calculated from $\dot{q}_1=0$ and
$q_2$ is calculated straightforwardly from the charge conservation equation. We
then minimise the value of $\dot{n}_{\rm i}$ along the curve $\dot{n}_{\rm
e}=0$.  All zero points are found by bisection to a tolerance level of
$10^{-20}$. We implemented the algorithm in Fortran 90 and compiled with
quadruple precision in order to avoid underflows in the very large terms that
are present in the dynamical equations.

We consider electric field strengths $E$ between $10^{-4}\,{\rm V\,m^{-1}}$ and
$10^3\,{\rm V\,m^{-1}}$ and a density of small particles relative to the gas
density $Z_1$ between $10^{-3}$ and $10^4$. The density of pebbles is fixed at a
10 times higher value, $Z_2 = 10 Z_1$. The radius of small dust is set to 1
$\mu$m and the radius of pebbles to 1 mm; both components are assumed spherical
with an internal density of $\rho_\bullet = 10^3\,{\rm kg\,m^{-3}}$. For each
value of $E$ and $Z_1$ we solve the charge equilibrium equations to obtain
$n_{\rm e}$, $n_{\rm i}$, $q_1$ and $q_2$.

\subsection{Results}

The equilibrium state is plotted in Figure \ref{f:charge}. We show the electron
and ion densities, the charge on the macroscopic pebbles and the current
densities of electrons and ions, all as functions of the electric field and the
density of small dust grains. We also indicate the critical electric fields for
electron heating (blue dotted line) and ion heating (red dotted line) as well as
the break down electric field to trigger lightning discharge (black dashed
lines), for the nominal value of \citet{DeschCuzzi2000} as well as for a 10
times lower value that may reflect the actual breakdown field strength [equation
(\ref{eq:EBD})]. Both ion and electron densities fall after reaching their
respective critical electric field strengths, due to the increase in collision
rates with dust. The ion density falls faster than the electron density as the
ions interact weakly with neutrals and hence are accelerated to very high speeds
that result in high collision rates with the dust particles. The electron
density falls much more slowly when the mass loading reaches $Z_1\sim100$
(corresponding to a total metallicity of $Z \sim 10^3$, see Figure
\ref{f:zeta_Z}). This is due to the release of secondary electrons as positrons
pass through lots of dust grains at those densities.

\subsection{Limiting values for electron and ion currents}

\cite{OkuzumiInutsuka2015} showed that the heating of ions causes the ion
current density to plateau out at
\begin{eqnarray}
  J_{{\rm i},\infty} &=& \frac{e \zeta n_{\rm n}}{\sigma_1 n_1} = \frac{e \zeta
  m_1}{Z_1 \sigma_1 m_n} \nonumber \\ &\approx& 1.1 \times 10^{-14}\,{\rm
  C\,m^{-2}\,s^{-1}} \left( \frac{Z_1}{100} \right)^{-1} \left( \frac{R_1}{\rm
  \mu m} \right) \, .
  \label{eq:Jiinf}
\end{eqnarray}
Here $\zeta=1.7 \times 10^{-17}\,{\rm s^{-1}}$ is the ionisation rate by
positrons from $^{26}$Al, valid when $Z \gg 1$ (see Figure \ref{f:zeta_Z}).
The limiting ion current density is in good agreement with Figure
\ref{f:charge}. For the electrons we obtain the limiting expression
\begin{eqnarray}
  J_{{\rm e},\infty} &=& \sqrt{\frac{\pi m_{\rm e}}{3 m_{\rm n}}} \frac{e \zeta
  n_{\rm n}}{\sigma_1 n_1} = \sqrt{\frac{\pi m_{\rm e}}{3 m_{\rm n}}} J_{{\rm
  i},\infty} \nonumber \\
  &\approx& 1.9 \times 10^{-16} \,{\rm C\,m^{-2}\,s^{-1}} \left(
  \frac{Z_1}{100} \right)^{-1} \left( \frac{R_1}{\rm \mu m} \right) \, .
\end{eqnarray}
This is also in excellent agreement with Figure \ref{f:charge} until $Z\approx
100$. Beyond this mass loading the electron production rate rises due to release
of secondary electrons. The electron current is nevertheless less important than
the ion current in setting the overall current density at all relevant values of
the mass loading. Importantly, the limiting values of the ion and electron
currents are independent of the gas density and hence the results presented in
Figure \ref{f:charge} are relatively independent of the distance from the star,
except for a small effect of the disc temperature on the conductivities.

\subsection{Charge flux}

Figure \ref{f:charge} shows that the charge on large pebbles, $q_2$, is positive
for high mass loading. This is due to the emission of secondary electrons as a
positron passes through a particle surface.  Pebbles are large enough to
completely stop positrons.  Therefore a positron will on the average pass
through $\lambda_2/\lambda_1$ small dust grains before it is absorbed by a
pebble, with $\lambda_1$ and $\lambda_2$ indicating the mean free path of the
positron relative to dust and pebbles. The relative production rate of secondary
electrons is
\begin{equation}
  f_{12} = \frac{N_{\rm s1}}{N_{\rm s2}} \frac{\lambda_2}{\lambda_1}
         = \frac{N_{\rm s1}}{N_{\rm s2}} \frac{A_1}{A_2} \, .
\end{equation}
Here $A_1/A_2$ is the total surface area of dust grains relative to the total
surface area of pebbles within a given volume. The total adsorption rate of
secondary electrons by the two dust components is also proportional to $A_i$,
Therefore, secondary electron release only yields net charging if $N_{\rm s1}$
is not equal to $N_{\rm s2}$. In Appendix \ref{s:secondary} we show that
$N_{\rm s1}=0.04$ and $N_{\rm s2}=0.08$. This asymmetry in the secondary
electron production yields net positive charging of the pebbles at the expense
of the dust grains. The resulting pebble charge can be very high, up to $10^5$
electron charges for strong electric fields. The charging of pebbles takes much
higher values when the charge flux from collisions with small dust is ignored
(as indicated by the black contour lines in the upper right plot of Figure
\ref{f:charge}).

\begin{figure}
  \begin{center}
    \includegraphics[width=0.9\linewidth]{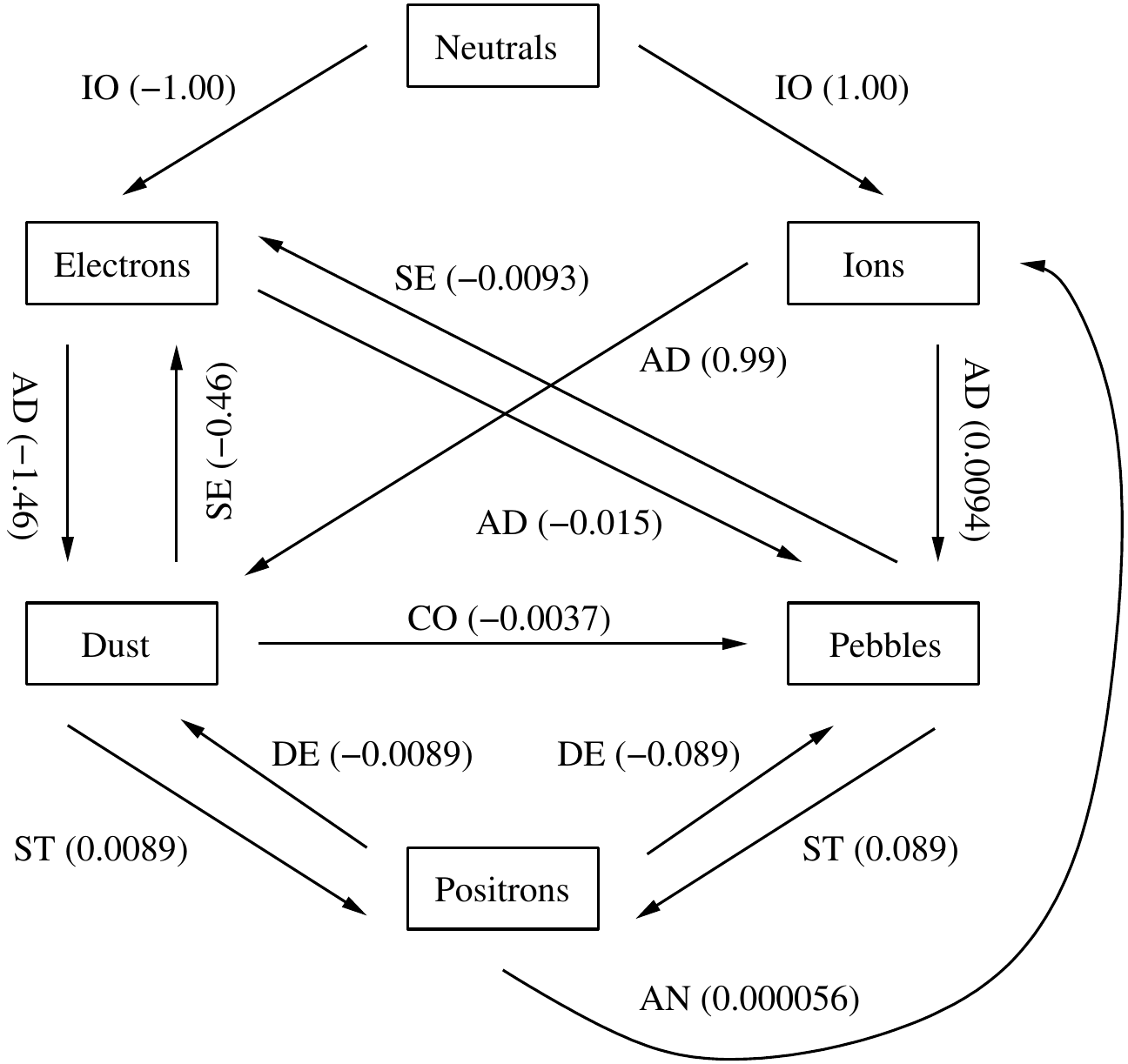}
  \end{center}
  \caption{Diagram of the charge fluxes between the five components neutrals,
  electrons, ions, dust, pebbles and positrons for $E=10\,{\rm V\,m^{-1}}$ and
  $Z_1=100$. The fluxes are normalised by the charge flux to dust and pebbles by
  ionisation. The abbreviations are: IO=ionisation, AD=adsorption, SE=secondary
  electrons, CO=dust collisions, DE=nuclear decay and positron emission,
  ST=stopping of positrons in solids, AN=annihilation of positrons in gas.}
  \label{f:diagram}
\end{figure}

In Figure \ref{f:diagram} we show the charge flux for the six components of the
charging model (including the neutrals), normalised to the charge flux to the
dust and pebble component by positron ionisation, for $Z_1=100$ and $E=10\,{\rm
V\,m^{-1}}$. Charging by
decay of $^{26}$Al (DE) and stopping of positrons (ST) are in close balance for
both dust and pebbles, with just a small fraction of positrons stopped by the
gas (AN). The production rate of secondary electrons (SE) by the dust is less
than 100 times (i.e., the ratio of total dust surface to total pebble surface)
that of the pebbles, due to the lower efficiency of secondary electron
production of the dust. This asymmetry of secondary electron production is
indeed the source of pebble charging, as the pebbles must charge positively to
pull in additional electrons that can compensate for the loss of secondary
electrons. Charge transfer in dust collisions (CO) is another source of pebble
flux on the pebbles; as is evident in Figure \ref{f:charge} this dust flux
reduces the pebble charge by approximately a factor ten at $(E,Z_1)=(10\,{\rm
V\,m^{-1}},100$).

\subsection{Drift current}

\begin{figure}
  \begin{center}
    \includegraphics[width=0.9\linewidth]{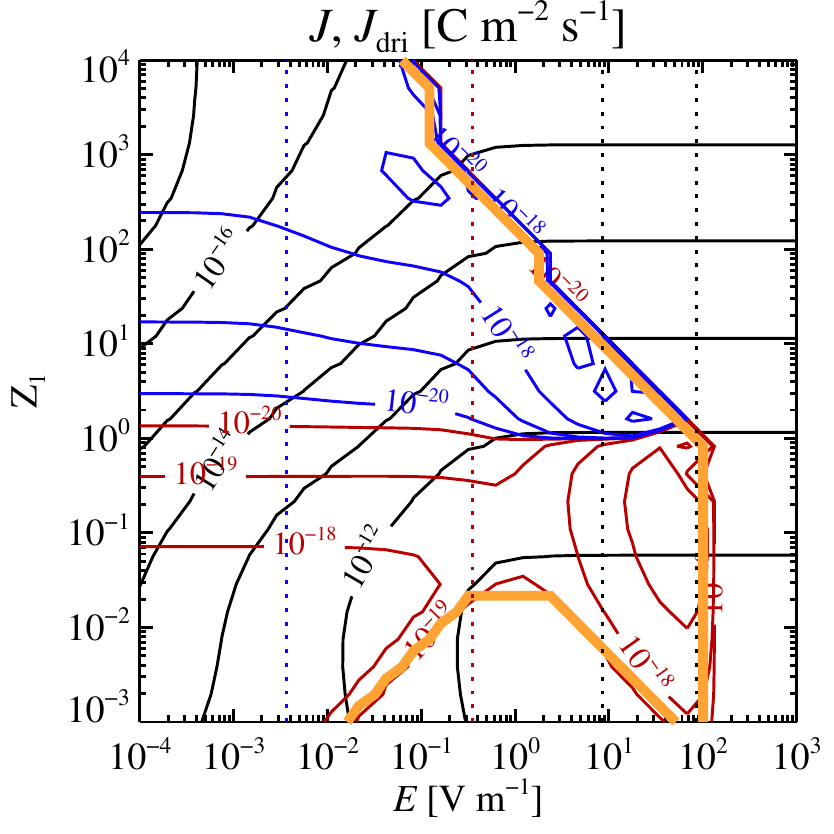}
  \end{center}
  \caption{The neutralising gas current $J$ (black contours) and the drift
  current $J_{\rm dri}$ (red for positive current and blue for negative), as a
  function of electric field $E$ and dust metallicity $Z_1$. The total current
  of the solids is separating to the left of the thick yellow zero-line and
  neutralising to the right of the zero-line. The drift current is at best 2-3
  orders of magnitude smaller than the neutralising gas current.}
  \label{f:pebble_current}
\end{figure}

The large, positive charging of the pebbles implies that spatially separating
pebbles from gas and dust will lead to the build-up of an electric field. The
question is whether the charge separation is continuously neutralised by the
electron/ion current or whether it builds up rapidly enough to necessitate
neutralisation by lightning discharge. We compare in Figure
\ref{f:pebble_current} the drift current to the gas current. We calculate
the drift current from
\begin{equation}
  J_{\rm dri}(E,Z_1) = n_1 e q_1 v_{1x} + n_2 e q_2 v_{2x} + n_{\rm i} e u_x -
  n_{\rm e} e u_x \, ,
\end{equation}
where $v_{1x}$, $v_{2x}$ and $u_x$ are the pebble, dust and gas velocities in
the radial direction.  These velocities result from both radial drift, as the
pressure gradient on the gas forces large pebbles to drift outwards at the cost
of small dust and gas drifting inwards, as well as the electric acceleration of
the dust and pebbles.  The equilibrium drift velocities are found by finding the
dynamical equilibrium of the gas velocity $(u_x,u_y)$ and the particle
velocities $(v_{1x},v_{1y})$, $(v_{2x},v_{2y})$ in the local frame corotating
with the protoplanetary disc at the Keplerian frequency $\varOmega$. This
equilibrium is given by the condition
\begin{eqnarray}
  \dot{u}_x = 0 &=& 2 \varOmega u_y - \frac{Z_1}{\tau_1} (u_x-v_{1x}) -
  \frac{Z_2}{\tau_2} (u_x-v_{2x}) - 2 \varOmega \Delta v \, , \label{eq:uxdot} \\
  \dot{u}_y = 0 &=& -\frac{1}{2} \varOmega u_x - \frac{Z_1}{\tau_1}
  (u_y-v_{1y}) - \frac{Z_2}{\tau_2} (u_y-v_{2y}) \, , \\
  \dot{v}_{1x} = 0 &=& 2 \varOmega v_{1y} - \frac{1}{\tau_1} (v_{1x}-u_x) +
  \frac{q_1}{m_1} E_x \, , \\
  \dot{v}_{1y} = 0 &=& -\frac{1}{2} \varOmega v_{1x} - \frac{1}{\tau_1}
  (v_{1y}-u_y) \, , \\
  \dot{v}_{2x} = 0 &=& 2 \varOmega v_{2y} - \frac{1}{\tau_1} (v_{2x}-u_x) +
  \frac{q_2}{m_2} E_x \, , \\
  \dot{v}_{2y} = 0 &=& -\frac{1}{2} \varOmega v_{2x} - \frac{1}{\tau_1}
  (v_{2y}-u_y) \, . \label{eq:vy2dot} 
\end{eqnarray}
Here $\tau_1$ and $\tau_2$ are the friction times of dust and pebbles,
respectively, and $\Delta v$ denotes the sub-Keplerian speed resulting from the
radial pressure gradient \citep{Nakagawa+etal1986}. In Figure
\ref{f:pebble_current} we overplot contours of the drift current on the
neutralising gas current. The drift current is separating (opposite sign as the
electric field) to the left of the zero-line neutralising (same sign as the
electric field) to the right of the zero-line. The zero-line of the drift
current is reached well before the breakdown value of the electric field at high
dust mass loadings. This implies that the mobility of the solids leads to
neutralisation of the charge separation well before lightning discharge can
occur.

\begin{figure}
  \begin{center}
    \includegraphics[width=0.9\linewidth]{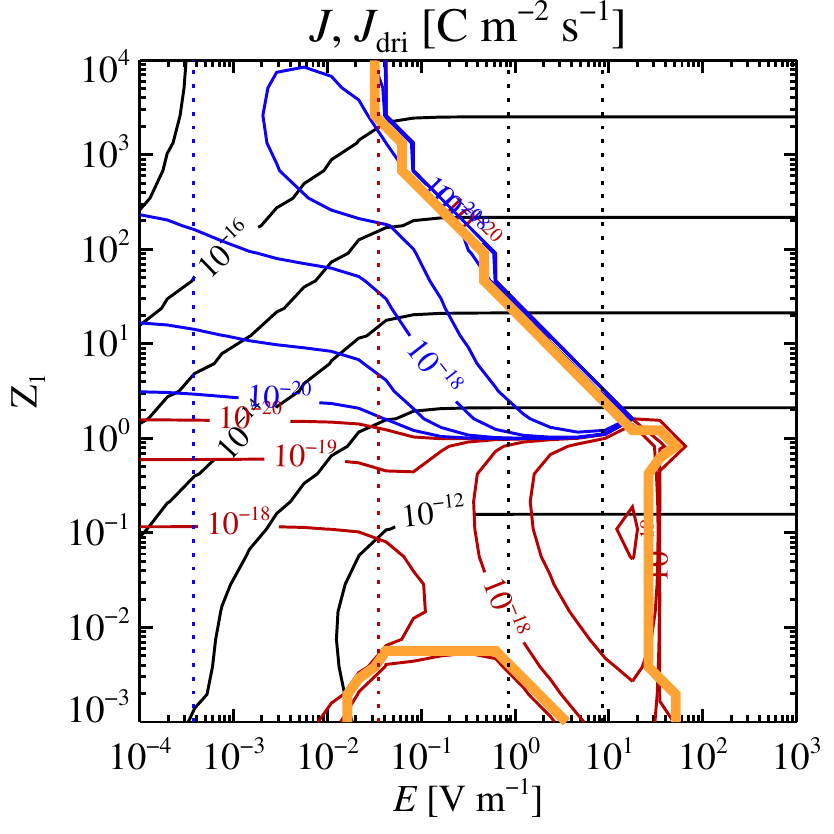}
  \end{center}
  \caption{Neutralising gas current and drift current for a model with reduced
  $^{26}$Al abundance (10\% of nominal value) and decreased gas density
  ($\rho_{\rm g}=10^{-8}\,{\rm kg\,m^{-3}}$). Both the neutralising gas current
  and the breakdown value of the electric field are a factor 10 lower than in
  Figure \ref{f:pebble_current}, but the drift current is still at least a
  factor 10 too small to compete with the neutralising current.}
  \label{f:pebble_current_lowbeta}
\end{figure}

In Figure \ref{f:pebble_current_lowbeta} we present a model where the abundance
of $^{26}$Al has been reduced by a factor 10 relative to the nominal value.
This leads to a corresponding decrease in the gas conductivity (see equation
\ref{eq:Jiinf}). We also reduced the gas density by a factor 10 to mimic
conditions at the late evolution stages of a protoplanetary disc after a few Myr
of gas accretion onto the star. The pebble charge is actually not reduced by the
decrease of the positron production, since the equilibrium is set by the
relative efficiency of secondary electron production between small dust grains
and macroscopic pebbles. The drift current is also relatively
unaffected by the change of gas density, since the reduction in the number
density of the solids is balanced by their higher mobility relative to the gas.
The lowered gas density decreases both the critical electric field strengths for
electron and ion heating as well as the breakdown electric field. The results
nevertheless show that the zero-drift current remains relatively unchanged,
although the lowered breakdown value of the electric field has approached the
zero-drift current by almost a factor 10.  A further reduction in the gas
conductivity and the breakdown electric field by an order of magnitude could
open a pathway to separating charges faster than neutralisation. However, such
low gas densities would imply particle densities much lower than those required
for chondrule formation.

In conclusion, it is very hard to obtain charge-separating radial drift currents
that are higher than the neutralising current. An additional concern about
chondrule formation by lightning in overdense filaments is the time-scale for
which these high densities can be sustained. Simulations of planetesimal
formation by the streaming instability demonstrate that the particle densities
can reach very high values of several thousand times the gas density during the
collapse \citep{Johansen+etal2015}. This phase is nevertheless short, on the
order of a few orbital periods, and hence can not lead to widespread thermal
processing of the particles involved in the collapse. Therefore we consider in
Section \ref{s:nuclear_battery} a potentially much power powerful source of
currents in protoplanetary discs, namely the flux of positrons emanating from
dense pebble regions.

\section{Adsorption rates}
\label{s:Kdi}

The cross section $\sigma_{dj}$ for collisions between a charged species $j$ and
a dust grain of type $d$ with charge $q_d$, radius $R_d$ and surface potential
$\phi_d=(4 \pi \epsilon_0)^{-1} q_d/R_d$ is given by \citep[see
e.g.][]{ShuklaMamun2002}
\begin{equation}
  \sigma_{dj} = \pi R_d^2 \left[ 1 - \frac{2 (\pm e) \phi_d}{m_j v_j^2} \right]
  \, .
\end{equation}
Here $m_j$ and $v_j$ are the mass and speed of the charged species and the
$\pm$ symbol refers to ions of charge $+e$ and electrons of charge $-e$,
respectively. The index $j$ can now be used to sum over a velocity
distribution. We denote the electron temperature by $T_{\rm e}$ and the ion
temperature by $T_{\rm i}$. These temperatures can obtain values different from
the thermal temperature $T$ due to heating by the electric field, see
discussion in Section \ref{s:eiheating}. For the electrons we assume that the
mean speed is negligible compared to the thermal speed
\citep{OkuzumiInutsuka2015,MoriOkuzumi2016}. The absorption rates of electrons
are therefore given by expressions that are independent of the mean speed,
\begin{eqnarray}
  K_{d{\rm e}} &=& \pi R_d^2 \sqrt{\frac{8 k_{\rm B} T_{\rm e}}{\pi m_{\rm e}}}
  \left( 1 + \frac{e \phi_d}{k_{\rm B} T_{\rm e}} \right) \quad {\rm
  for\,\,\phi_{\it d}>0} \, , \\
  K_{d{\rm e}} &=& \pi R_d^2 \sqrt{\frac{8 k_{\rm B} T_{\rm e}}{\pi m_{\rm e}}}
  \exp \left( \frac{e \phi_d}{k_{\rm B} T_{\rm e}} \right) \quad {\rm
  for\,\,\phi_{\it d}<0} \, .
\end{eqnarray}
The adsorption rate of ions must be calculated by considering the separate
flux contributions from the ion temperature $T_{\rm i}$ and the ion speed along
the electric field, $\langle \vc{v}_{\rm i} \rangle$. The adsorption rate is
calculated through the integral
\begin{eqnarray}
K_{d{\rm i}}  &\equiv& \pi a^2  \int d^3 v_{\rm i} \,\, v_{\rm i}
\left(1-\frac{2e\phi_d}{m_{\rm i} v_{\rm i}^2}\right) f_{\rm i}({\bm v}_{\rm i})
\nonumber \\
&=&  \pi a^2  \int_{v_0}^\infty dv_{\rm i} \,\, 4\pi v_{\rm i}^3
\left(1-\frac{2e\phi_d}{m_{\rm i} v_{\rm i}^2}\right) \nonumber \\
&& \qquad \qquad \qquad \qquad \times \int_{\cos\theta=-1}^{\cos\theta=1}
d(\cos\theta) f_{\rm i} (v_{\rm i}, \theta) \, .
\end{eqnarray}
We then insert a Maxwellian distribution function that includes a mean velocity
$\langle \vc{v}_{\rm i} \rangle$,
\begin{eqnarray}
f_{\rm i}({\bm v}_{\rm i},\theta) &=& \left(\frac{m_{\rm i}}{2\pi k_{\rm B} T_{\rm
i}}\right)^{3/2} \exp\left[-\frac{m_{\rm i}}{2k_{\rm B} T_{\rm i}}({\bm v}_{\rm
i}-\langle{\bm v}_{\rm i}\rangle)^2\right] \nonumber \\
&=& \left(\frac{m_{\rm i}}{2\pi k_{\rm B} T_{\rm i}}\right)^{3/2}
\exp\left[-\frac{m_{\rm i}}{2k_{\rm B} T_{\rm i}}(v_{\rm i}^2 + |\langle{\bm v}_{\rm
i}\rangle|^2 - 2v_{\rm i}|\langle{\bm v}_{\rm i}\rangle|\cos\theta )\right]
\, . \nonumber \\
\end{eqnarray}
Here $v_{\rm i} = |{\bm v}_{\rm i}|$ is the magnitude of ${\bm v}_{\rm i}$ and
$\theta$ is the angle between ${\bm v}_{\rm i}$ and $\langle{\bm v}_{\rm
i}\rangle$. We denote by $v_0$ the collision speed below which the effective
collision cross section $\pi R_d^2 [1-2e\phi_d/(m_{\rm i} v_{\rm i}^2)]$ is
zero. That gives
\begin{eqnarray}
  v_0 &=& \sqrt{\frac{2e\phi_d}{m_{\rm i}}} \qquad {\rm for}\,\,\phi_d>0 \, , \\
  v_0 &=& 0 \qquad \qquad \quad {\rm for}\,\, \phi_d<0 \, .
\end{eqnarray}
For the repulsion case with $\phi_d > 0$ we obtain
\begin{eqnarray}
K_{d{\rm i}} &=& \pi R_d^2\Biggl\{  \sqrt{\frac{k_{\rm B}T}{2\pi m_{\rm i}}}
\left[ \left(1+\frac{v_0}{|\langle\bm v_{\rm
i}\rangle|}\right)\exp\left(-\frac{m_{\rm i}(v_0-|\langle\bm v_{\rm
i}\rangle|)^2}{2k_{\rm B}T_{\rm i}}\right) + \right.  \nonumber \\
&& \left. \left(1-\frac{v_0}{|\langle\bm v_{\rm
i}\rangle|}\right)\exp\left(-\frac{m_{\rm i}(v_0+|\langle\bm v_{\rm
i}\rangle|)^2}{2k_{\rm B}T_{\rm i}}\right) \right]  \Biggr. + \nonumber \\
&& \Biggl.  \frac{|\langle\bm v_{\rm i}\rangle|}{2}\left(1+\frac{k_{\rm B}T_{\rm
i}-2e\phi_d}{m_{\rm i} |\langle\bm v_{\rm i}\rangle|^2}\right) \left[ {\rm
erf}\left(\sqrt{\frac{m_{\rm i}}{2k_{\rm B}T_{\rm i}}}(|\langle\bm v_{\rm
i}\rangle|-v_0)\right) + \right. \nonumber \\
&& \left. {\rm erf}\left(\sqrt{\frac{m_{\rm i}}{2k_{\rm B}T_{\rm i}}}(|\langle\bm v_{\rm
i}\rangle|+v_0)\right) \right] \,\, \Biggr\} \, .
\end{eqnarray}
For the attraction case with $\phi_d < 0$ we obtain an expression already given
in \cite{ShuklaMamun2002},
\begin{eqnarray}
K_{d{\rm i}} &=& \pi R_d^2 \left\{ \sqrt{\frac{2k_{\rm B}T}{\pi m_{\rm i}}}
\exp\left(-\frac{m_{\rm i} |\langle\bm v_{\rm i}\rangle|^2}{2k_{\rm B}T_{\rm i}}
\right) \right. \nonumber \\
&& \left. + |\langle\bm v_{\rm i}\rangle|\left(1+\frac{k_{\rm B}T_{\rm
i}-2e\phi_d}{m_{\rm i} |\langle\bm v_{\rm i}\rangle|^2}\right) {\rm
erf}\left(\sqrt{\frac{m_{\rm i}}{2k_{\rm B}T_{\rm i}}}|\langle\bm v_{\rm
i}\rangle|\right) \right \} \, .
\end{eqnarray}
Note that the repulsion and accretion expressions for $K_{d{\rm i}}$ are the
same for $\phi_d=v_0=0$.

\subsection{Dust adsorption}

The adsorption of small dust grains by large pebbles is given by the expression
\begin{equation}
  K_{12} = \sigma_2 v_{12} \left( 1 - \frac{q_1 e \phi_2}{E_1} \right) \, .
\end{equation}
Here $v_{12}$ is the relative drift speed between dust grains and pebbles,
calculated from the drag force equilibrium of Equations
\ref{eq:uxdot}--\ref{eq:vy2dot}, and $E_1$ is the kinetic energy
\begin{equation}
  E_1 = \frac{1}{2} m_1 v_{12}^2 \, .
\end{equation}

\section{Secondary electrons}
\label{s:secondary}

Positrons released by $^{26}$Al in regions of very high particle density are
primarily stopped by other particles. The passage of a positron through the
surface of a solid particle is accompanied by the emission of $N_{\rm s}$
secondary electrons. This leads to a net negative charging of dust particles,
since a small fraction of the positrons are absorbed by the gas. The secondary
electrons also contribute to the conductivity of the gas. In this appendix we
calculate the yield of secondary electrons as a function of the size of the
particle that the positron passes through. The secondary electron yield is an
important parameter in the charge equilibrium model presented in the main paper.
\begin{figure}
  \includegraphics[width=\linewidth]{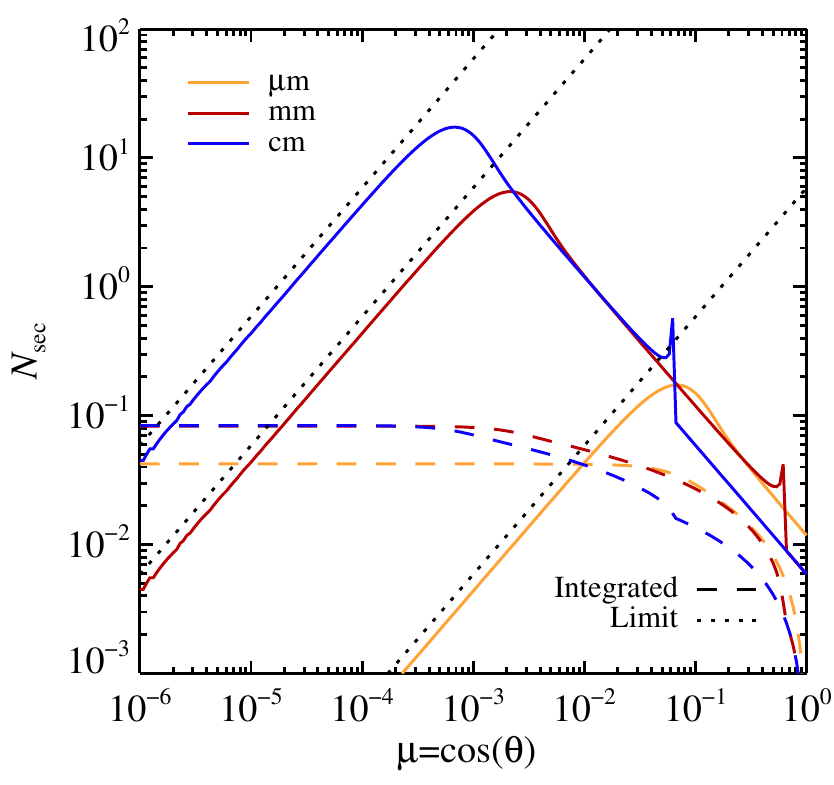}
  \caption{Secondary electron production versus the incident angle
  $\mu=\cos(\theta)$ for particles of radii $\mu$m, mm and cm. The secondary
  electron production rises with decreasing $\mu$ and experiences a jump at the
  angle when the positron penetrates all the way through the solid. Finally,
  the secondary electron production falls when the pathway through the particle
  is short at very low $\mu$ (dotted line). The integrated yield (dashed lines)
  is about 0.08 for mm and cm solids and 0.04 for $\mu$m solids.}
  \label{f:Nsec_mu}
\end{figure}

The production of secondary electrons by a positron experiencing energy loss
$\de E_\beta$ in the interval between $\chi$ and $\chi+\de\chi$ along its
penetration path through a solid particle can be written according to
\cite{LinJoy2005} as
\begin{equation}
  \de N_{\rm sec} = - \frac{\de E_\beta}{\de \chi} \frac{1}{\epsilon} 0.5
  \exp(-\alpha \chi_\perp) \de \chi \, .
  \label{eq:dNsec}
\end{equation}
Here $\epsilon$ is the ionisation potential of electrons inside the material,
$\alpha$ is the inverse stopping length of the released electrons and
$\chi_\perp$ is the shortest distance to the surface of the particle. The
particle sizes we consider are much larger than the stopping length of the
electrons and hence the value of $\alpha$ obtained from experiments with flat
surfaces can be applied to curved particles as well. The perpendicular distance
$\chi_\perp$ can be written as function of $\chi$ for a positron with angle of
incidence $\theta$ relative to the normal of the surface as
\begin{equation}
  \chi_\perp = \sqrt{R^2 + \chi^2 - 2 R \chi \cos(\theta)} \, .
\end{equation}
We use the parameters $\epsilon=90$ eV and $\lambda=1/\alpha=2.7$ nm relevant
for pure silicon \citep{LinJoy2005}. The energy loss rate is calculated from
the Bethe-Bloch equation modified for positrons \citep[e.g.][chapter
31]{Olive+etal2014},
\begin{equation}
  \frac{\de E}{\de \chi} = -\frac{1}{2} K \rho_{\rm i} \frac{Z_{\rm i}}{A_{\rm
  i}} \frac{1}{\beta^2} \left\{ \ln \frac{m_{\rm e} c^2 \beta^2 \gamma^2
  [m_{\rm e} c^2 (\gamma-1)]}{2 I_{\rm i}^2} + F \right\} \, .
\end{equation}
Here the constant $K=4 \pi N_{\rm A}r_{\rm e}^2 m_{\rm e} c^2$, with $N_{\rm
A}$ denoting Avogadro's number, $r_{\rm e}$ the classical electron radius,
$m_{\rm e}$ the electron mass and $c$ the speed of light.  The material
is characterised by the density $\rho_{\rm i}$, nuclear charge $Z_{\rm i}$ and
atomic number $A_{\rm i}$. The positron energy is defined through $\beta = v/c$
and $\gamma = 1/\sqrt{1-\beta^2}$. The function $F$ is specific to positrons
and is defined as
\begin{equation}
  F = 2 \ln 2 - \frac{\beta^2}{12} \left[ 23 + \frac{14}{\gamma+1} +
  \frac{10}{(\gamma+1)^2} + \frac{4}{(\gamma+1)^3} \right] \, .
\end{equation}
We calculate the number of secondary electrons produced as a function of $\mu =
\cos(\theta)$ for particles of $\mu$m, mm and cm sizes. Secondary electrons are
mainly produced within a few times the mean free path $\lambda$ from the
surface. We therefore only integrate \Eq{eq:dNsec} at distances within $10
\lambda$ from the surface, in order to speed up the integration. We use an
energy of $E_0=0.66$ MeV for the positron. The results are shown in
\Fig{f:Nsec_mu}. The yield of secondary electrons increases initially with
decreasing $\mu$, as the positron moves closer to the surface.  The yield
experiences a sudden increase at an angle where the positron passes all the way
through the particle. Finally, the yield falls along a characteristic curve
\begin{equation}
  N_{\rm sec}=\frac{2 R \mu}{\chi_{\rm s}} \frac{0.5 E_0}{\epsilon}
\end{equation}
for very low values of $\mu$. Here $2 R \mu$ is the passage length through the
particle and $\chi_{\rm s}$ is the total stopping length in the material (1 mm
for silicate particles). The integrated yield gives the total number of
secondary electrons through
\begin{equation}
  N_{\rm sec} = (2 \pi)^{-1} \int_0^{2 \pi} \int_0^{\pi/2} \de N_{\rm sec}
  \sin \theta \de \theta \de \phi = \int_0^1 \de N_{\rm sec} \de \mu \, .
\end{equation}
This gives an integrated yield of $N_{\rm sec}=0.08$ for particles of mm and cm
sizes and $N_{\rm sec}=0.04$ for $\mu$m-sized particles.

\end{document}